\documentclass[article]{elsarticle}
\usepackage{hyperref}
\usepackage{amsfonts,amsmath,graphicx,color,epsfig}
\usepackage{natbib}

\usepackage{colortbl}

\usepackage{multirow}% http://ctan.org/pkg/multirow
\usepackage{hhline}% http://ctan.org/pkg/hhline
\newcommand{\ie}{{\it i.e. }}

\newcommand{\bR}{{\mathbb{R}}}
\newcommand{\bP}{{\mathbb{P}}}
\newcommand{\bE}{{\mathbb{E}}}
\newcommand{\lck}{\left\lbrack}
\newcommand{\rck}{\right\rbrack}
\newcommand{\lce}{\left\lbrace}
\newcommand{\rce}{\right\rbrace}
\newcommand{\lp}{\left(}
\newcommand{\rp}{\right)}

\newcommand{\dd}{\, \text{d}}

\newcommand{\un}{\mathbb{I}}
\newcommand{\Id}{\mbox{I} \!\!\mbox{I}}
\newcommand{\Un}{{\bf 1}}

\newcommand{\var}{\mathbb{V}\mathrm{ar}}
\newcommand{\cov}{\mathbb{C}\mathrm{ov}}

\newtheorem{theorem}{Theorem}[section]
\newtheorem{lemma}[theorem]{Lemma}
\newtheorem{proposition}[theorem]{Proposition}

\newcommand{\finpreuve}{\hfill $\Box$ \\}

\definecolor{grey}{rgb}{0.5,0.5,0.5}

\begin{document}

\begin{frontmatter}

%\title{Defining a variographic approach from the characteristics of a point process to predict the local intensity of partially observed data}
\title{Predicting the intensity of partially observed data from
a revisited kriging for point processes}

\author[mymainaddress,myfourthaddress]{Edith Gabriel\corref{mycorrespondingauthor}}
\cortext[mycorrespondingauthor]{Corresponding author}
 \ead{edith.gabriel@univ-avignon.fr}

\author[mymainaddress]{Florent Bonneu}
\author[mysecondaryaddress,myfourthaddress]{Pascal Monestiez}
\author[mythirdaddress]{Jo{\"e}l Chad{\oe}uf}

\address[mymainaddress]{Avignon University, LMA EA 2151, 84000 Avignon, France}
\address[mysecondaryaddress]{INRA, BioSP, 84000 Avignon, France}
\address[mythirdaddress]{INRA, Statistics, GAFL, UR 1052, 84000 Avignon, France}
\address[myfourthaddress]{Zone Atelier Plaine \& Val de S{\`e}vre, CEBC-CNRS, 79360 Villiers-en-Bois, France }

\begin{abstract}
We consider a stationary and isotropic spatial point process, whose a realisation is observed within a large window.
In order to predict its local intensity, we propose to define the first-
and second-order characteristics of a random field, defined as the regularized counting process, from the ones of the point process
and to interpolate the intensity by using a revisited kriging of the regularized process.
\end{abstract}

\begin{keyword}
Intensity estimation; Point process; Prediction; Spatial statistics.
%\texttt{elsarticle.cls}\sep \LaTeX\sep Elsevier \sep template
%\MSC[2010] 00-01\sep  99-00
\end{keyword}

\end{frontmatter}

%\linenumbers

\section{Introduction}

When estimating the intensity of a point process, we observe the full point
pattern within a window % which is the set of observed areas
and we want to know its local changes over a given mesh.
This issue has been addressed in several ways: kernel smoothing, see \cite{silverman1986}, \cite{guan2008} in presence of covariates, and
\cite{vanlieshout2012} for a general class of weight function estimators that encompasses both kernel and tessellation based
estimators; or parametrical methods; see for instance
\cite{illian2008} for a review.
A recurrent and remaining question in these approaches is which bandwidth/mesh
should we use? This has been addressed by using cross-validation \cite{hardle1991}
or double kernel \cite{devroye1989}.

In contrast to the previous methods which look at the intensity changes
inside the observation window, our main interest lies in predicting
the intensity outside the observation window, all the more when it is
not connected as it frequently happens when sampling in plant ecology.
To predict the intensity we could use
\cite{tscheschel2006}'s reconstruction method based on the first- and second-order
characteristics of the point process.
Once the empirical point pattern predicted within a given
window, one can get the intensity by kernel smoothing.
As it is a simulation-based method, it requires long computation times, especially
when the prediction window is large and/or the point process is complex.
As alternative method, few authors model the point pattern by a point process with the intensity driven by
a stationary random field. In \cite{diggle2007} and \cite{diggle2013},
the approach is heavily based on a complete modelling and considers a log-Gaussian model.
The parameter estimation, the intensity estimation and
its prediction outside the observation window are obtained using a bayesian framework.
The method developed in \cite{monestiez2006} and \cite{bellier2013} is close to
classical geostatistics. Basically, it consists in counting the number of points
within some grid cells, computing the related empirical variogram and theoretically
relating it to the one obtained from the random field driving the intensity. Then,
the variogram is fitted and kriging is used to predict the intensity.
 Its advantage is that the estimation is only based on its first- and second-order moments so that the model does not need to be fully specified.
While this approach
requires less hypotheses, the model remains constrained within the class of Cox processes. Moreover, the mesh size is arbitrary defined.

\bigskip

Here, we consider a similar approach as it does not require
model specification, but it addresses a larger class of point processes and optimises
the scale of investigation. The local intensity of a stationary and isotropic point process can be written as $\lambda + Y(x)$, where $\lambda$ is the
mean of the random field and $Y(x)$ is a centered random field.
The number of points within a Borel set $B$ is then
\begin{equation}
\label{eq:geostatdecomp}
\lambda \nu(B)+\int_B Y(x) \dd x + \eta,
\end{equation}
\ie the sum of the global mean, the local intensity variation and an error related
to the difference between the observations and the local intensity, respectively.
Equation~(\ref{eq:geostatdecomp}) is very similar to the geostatistical
decomposition.
%Thus, we propose to define the first- and second-order characteristics of the random field from the ones of the point process.
Thus, we propose to interpolate the local intensity by kriging,
where the kriging weights depend on the local structure of the point process.
Hence, our method uses all the data to locally predict at a given point,
which it is not the case of most of kernel methods, and it also uses the
information at fine scale of the point process, which it is not the case in
geostatistical approaches. Furthermore, it does not require a specific model
but only (an estimation of) the first- and second-order characteristics of
the point process.

\bigskip

In Section~\ref{sec:ppvsgeostat} we define a random field
of point counts on grid cells and we link up the mean and variogram of this random
field to the intensity and pair correlation function of the point process. The kriging
weights, the related interpolator and its properties are presented in
Section~\ref{sec:kriging} as well as the optimal mesh of the interpolation grid. In Section~\ref{sec:real} we use our kriging interpolator
to estimate and predict the intensity of Montagu's Harriers' nest locations in a
region of France. % and compare these results with those obtained by a kernel estimator.
In Section~\ref{sec:experiments}, we discuss the influence
of the mesh $B$ and the rate and shape of unobserved areas on the statistical properties of
our kriging interpolator from numerical results.

\section{Linking up characteristics of two theories}
\label{sec:ppvsgeostat}

Let $\Phi$ be a stationary and isotropic point process and $B$ a Borel set centered at $0$.
%, obtained by a weak dependent process with a parameter driven by a stationary random field at a larger scale (e.g. Poisson process versus Cox process).
Following the notations in \cite{stoyan1996}, a realisation of $\Phi$
within a window $S$ will be denoted by $\Phi_S$ and the random
counting measure for a Borel set $B$ by $\Phi(B)$.

\bigskip

In our context, data are defined as informative point locations (the
realisation of the point process $\Phi$)
while the geostatistical calculations (kriging) need to be carried out over the
values of a random field $Z$ observed at several sampling locations, grid cell centers
for example.
Thus, we must regularize our process over
a compact. This consists in defining $Z(x)$ by the count
of the point process over the grid cell $B$ centered at $x$
%of  a point process $\Phi$ evaluated in a grid cell $B$ centered at $x$,
\ie $Z(x) = \Phi(x \oplus B)$.
% Then we need to consider its characteristics $m^*$ and $\gamma^*$.

\subsection{About geostatistics}
Let $Z(x)$ be a real valued random field.
Its first-order characteristic is
the mean value function: $\bE \lck Z(x) \rck = m(x)$,
its second-order characteristics are classically described in geostatistics \citep{matheron1962,
matheron1963}
by the (semi)-variogram, \ie the mean squared
difference at distance $h$: $\gamma(h) = \frac{1}{2} \bE \lck \left(Z(x) -
Z(x+h)\right)^2 \rck$.
For a stationary and isotropic random field, we have
\begin{eqnarray}\label{eq:chargeostat}
  \bE \lck Z(x) \rck & = & m, \nonumber \\
  \gamma(h) & = & \sigma^2 - \cov(Z(x),Z(x+h)),
\end{eqnarray}
where $\sigma^2$ is the field variance and $\cov(Z(x),Z(x+h))$ is the auto-covariance of the random field.

\medskip

We can predict the value $Z(x_o)$ at the unsampled location $x_o$ by using
the best linear unbiased predictor, so-called kriging interpolator:
$\widehat Z(x_o) = \mu^T z$, where $z = \lce Z(x_i) \rce_{i=1,\dots,n}$ is the observations vector of the random field and $\mu$ is the $n$-vector of weights.
In the case of ordinary kriging \cite{cressie1993}, which is of interest here since
the mean value of the random field is unknown, we have
\begin{equation}
\label{eq:ko}
\mu = C^{-1} C_o + \frac{1- \Un^T C^{-1} C_o}{\Un^T C^{-1} \Un} C^{-1} \Un,
\end{equation}
where $C = \lce\cov \big(Z(x_i),Z(x_j) \big) \rce_{i,j=1,\dots,n}$ is the % $n \times n$
 covariance matrix
between the observations, $C_o = \lce \cov \big( Z(x_i),Z(x_o) \big) \rce_{i=1,\dots,n}$ is the covariance vector
between the observations and $Z(x_o)$ and $\Un$ is the $n$-vector of 1 (see e.g. \cite{cressie1993,wackernagel2003}).

\subsection{About point processes}
Let $\Phi$ be a point process defined in $\bR^2$ and observed in $S$.
Its first- and second-order characteristics
are described through its intensity
$\lambda$ and the Ripley's $K$-function or the pair correlation function $g$:
\begin{eqnarray}\label{eq:charpp}
\lambda & = & \frac{\bE \lck \Phi(S) \rck}{\nu(S)}, \\
K^*(r) & = & \dfrac{1}{\lambda} \bE \lck \Phi(b(0,r)) - 1 | 0 \in \Phi \rck, \\
g(r) & = & \dfrac{1}{2 \pi r} \dfrac{\partial K^*(r)}{\partial r},
\end{eqnarray}
where $\nu(S)$ is the area of $S$ and $b(0,r)$ is the disc centered at $0$,
with radius $r$.
The intensity $\lambda$ is thus the expected number of points per unit area,
$\lambda K^*(r)$ is the mean number of points in a circle of radius $r$ centered at a
typical point of the point process, whereas $g(r)$ measures how
$K^*$ changes with $r$. See for instance \cite{stoyan1996} for a review about the theory
of point processes.
\begin{lemma}
\label{lem:pp}
Let $\Phi$ be a point process with intensity $\lambda$
and $B$, $D$ two Borel sets.
 Then,
\begin{enumerate}
  \item If $\nu(B), \nu(D) \to 0$, then $\bP \lck \lce \Phi(B) = 1 \rce \cap \lce \Phi(D) = 1 \rce \rck
  = \lambda \nu(B \cap D) + \lambda^2 \int_{B \times D} g(x-y) \dd x \dd y
  + o\lp \nu(B \cup D) \rp$,

  \item $\bE \lck \Phi^2(B) \rck = \lambda \nu(B)
  + \lambda^2 \int_{B \times B} g(x-y) \dd x \dd y$,

  \item $\var (\Phi(B)) = \lambda \nu(B)
  + \lambda^2 \int_{B \times B} \left( g(x-y) - 1 \right) \dd x \dd y$,

  \item If $B \cap D = \emptyset$, then $\cov \left( \Phi(B), \Phi(D) \right) =
  \lambda^2 \int_{B \times D} \left( g(x-y) - 1 \right) \dd x \dd y$.
\end{enumerate}
\end{lemma}

\noindent The proofs are in \ref{app:A}.

\subsection{Linking up}
\label{sec:link}
From the first- and second-order moments defined in the previous sections,
we can link up the characteristics of the point process $\Phi$ to the ones of the
random field of point counts $Z$.
Because of the stationary assumption it can also be related to the
auto-covariance function (Equation~(\ref{eq:chargeostat})), thus in the following we
shall consider the latter.
\begin{proposition}
\label{prop:ppvsgeostat}
For the count random field defined by $\Phi(B)$, where $B$ is a given Borel set, we have:
\begin{enumerate}
  \item $m = \lambda \nu(B)$,
  \item For $B$ and $D$ two regularization blocks, $B_D = B \backslash D$,
  $D_B=D \backslash B$,
  \begin{eqnarray*}
  2 \gamma(B,D) &=& \lambda \lp \nu(B_D) + \nu(D_B) \rp
   + \lambda^2 \lp \int_{B_D \times B_D} g(x-y) \dd x \dd y \right. \\
  && \left. + \int_{D_B \times D_B} g(x-y)
   \dd x \dd y - 2 \int_{B_D \times D_B} g(x-y) \dd x \dd y \rp.
   \end{eqnarray*}
   \item If $B$ and $D$ are  centered at points
   with a distance $r$, then for $\nu(B) = \nu(D) \rightarrow 0$
    \begin{equation}
   \label{eq:cov}
    \cov \left( \Phi(B), \Phi(D) \right) \approx \lambda \nu(B) \Big(
    \un_{\lce B=D \rce}  + \lambda \nu(B) \big( g(r) - 1 \big) \Big).
    \end{equation}
\end{enumerate}
\end{proposition}
The proof of Proposition~\ref{prop:ppvsgeostat} is straightforward from
Lemma~\ref{lem:pp} and from the approximation %$B$, $D$ are centered at points with distance $r$.
$\bP \lck \lce \Phi(B) = 1 \rce \cap \lce \Phi(D) =1 \rce \rck \approx  \lambda^2 \nu(B) \nu(D) g(r)$~(see \ref{app:B}).

\section{Revisited kriging for point processes}
\label{sec:kriging}

We want to interpolate the intensity of the point process given its realisation  within an observation window $S_{obs}$, $\lambda(x | \Phi_{S_{obs}})$. Hence, we use the relation between point processes and geostatistics (section~\ref{sec:link}) and approximate the point process by the counting process within a grid of elementary cell $B$.

For sake of clarity, in the following we denote by $S$ the region of interest so that $S_{unobs}$ define the complementary of $S_{obs}$ within $S$. We consider a regular grid superimposed on $S$ with a square-mesh.
We denote by $B$ an elementary square centered at 0, $B_i = x_i \oplus B$ the elementary square
centered at $x_i$ such that $B_i \cap B_j = \emptyset$,
and $n$ (resp. $n_{obs}$) the number of grid cell centers lying in $S$ (resp. $S_{obs}$).
% Thus $S = \cup_{i=1}^n B_i$, $n = \frac{\nu(S)}{\nu(B)}$ and $n_{obs}=\frac{\nu(S_{obs})}{\nu(B)}$.

\subsection{Defining the interpolator}
According to the classical geostatistical method defined in
Section~\ref{sec:ppvsgeostat}, the kriging interpolator of the local intensity at $x_o$,
$\lambda(x_o | \Phi_{S_{obs}})$, should be written as
$$\mu^T \big(\lambda(x_1 | \Phi_{S_{obs}}), \dots,
\lambda(x_{n_{obs}} | \Phi_{S_{obs}}) \big),$$
for some well-chosen kriging weights $\mu$ where $x_i$, $i=1,\dots,n_{obs}$ correspond to data sample locations, i.e. here to the cell centers of $S_{obs}$.
Note that in our case we cannot observe the
local intensity at $x_i$, thus we can estimate it by
$\dfrac{\Phi(B_i)}{\nu(B)}$.
Furthermore because of the cell-point relation, we cannot have an exact interpolation of the local intensity.
\begin{proposition}\label{prop:interp}
Given the elementary square $B$, the interpolator at $x_o$ defined by
\begin{equation}\label{eq:interp}
\widehat \lambda(x_o | \Phi_{S_{obs}}) = \sum_{x_i \in S_{obs}} \mu_i \dfrac{\Phi(B_i)}{\nu(B)},
\end{equation}
where $\mu = (\mu_1, \dots, \mu_{n_{obs}}) =  C^{-1} C_o
+ \dfrac{1- \Un^T C^{-1} C_o}{\Un^T C^{-1} \Un} C^{-1} \Un$,
is the best linear unbiased predictor (BLUP) of $\dfrac{\Phi(B_0)}{\nu(B)}$ and the
asymptotically BLUP of $\lambda(x_o | \Phi_{S_{obs}})$.

\noindent
The weights depend on
\begin{itemize}
  \item the covariance matrix $C = \lambda \nu(B) \Id
  + \lambda^2 \nu^2(B) (G-1)$,

  where $G = \lce g_{ij} \rce_{i,j=1,\dots,n_{obs}}$, with $g_{ij} =  \frac{1}{\nu^2(B)}
  \int_{B \times B} g(x_i-x_j+u-v) \dd u \dd v $,
  and $\Id$ is the $n_{obs} \times n_{obs}$-identity matrix,
  \item the covariance vector $C_o = \lambda \nu(B) \un_{x_o}
  + \lambda^2 \nu^2(B) (G_o-1)$,

  where $G_o = \lce g_{io} \rce_{i=1,\dots,n_{obs}}$,
  and $\un_{x_o}$ is the $n_{obs}$-vector with zero values and
  one term equals to one where $x_o = x_i$ (which only happens
  in estimation).
\end{itemize}
\end{proposition}
\textbf{Proof:}
At the scale of $B$, the kriging weights such that $\widehat \lambda(x_o | \Phi_{S_{obs}})$ is a BLUP of $\dfrac{\Phi(B_o)}{\nu(B)}$ are given by the ordinary kriging
equations \cite{chiles2012}.

At a finer scale we have that $\bE \lck \dfrac{\Phi(B)}{\nu(B)} \rck$ tends to $\lambda(x)$ when $\nu(B)$ tends to 0 (as $\lambda(x)$ is assumed to be continuous). Thus we propose to interpolate $\lambda(x_o | \Phi_{S_{obs}})$ by using $\widehat \lambda(x_o | \Phi_{S_{obs}}) = \sum_{x_i \in S_{obs}} \mu_i \dfrac{\Phi(B_i)}{\nu(B)}$, with the constraint $\sum_{i=1}^{n_{obs}} \mu_i =1$.
Minimising the variance error $\var \lp \widehat \lambda(x_o | \Phi_{S_{obs}})
- \lambda(x_o | \Phi_{S_{obs}}) \rp$ under this constraint
 and using Equation~(\ref{eq:cov}) lead to the following kriging weights:
$$\mu =  \nu(B) C^{-1} \widetilde C_o
+ \dfrac{1- \nu(B) \Un^T C^{-1} \widetilde C_o}{\Un^T C^{-1} \Un} C^{-1} \Un,$$
where $\widetilde C_o = \lce \cov \Big( \Phi(B_i) , \lambda \big(x_o | \Phi_{S_{obs}} \big)
\Big) \rce_{i=1,\dots,n_{obs}}$.

\medskip

\noindent
To get $\cov \Big( \Phi(B_i) , \lambda \big(x_o | \Phi_{S_{obs}} \big) \Big)$, note that
\begin{itemize}
  \item for $x_i \neq x_o$, we have
\begin{eqnarray*}
  \bE \lck \Phi(B_o) \Phi(B_i) \rck &=& \bE \lck \Phi(B_i) \bE \lck
  \Phi(B_o) | \Phi(B_i) \rck \rck \\
  &=&  \bE \lck \Phi(B_i) \nu(B) \lambda(x_o |
  \Phi(B_i)) \rck \\
  & = &  \bE \lck \Phi(B_i) \nu(B) \bE \lck \lambda(x_o | \Phi_{S_{obs}} \rck |
  \Phi(B_i) \rck \\
  & =&   \nu(B)  \bE \lck \bE \lck \Phi(B_i) \lambda(x_o | \Phi_{S_{obs}})
   \rck | \Phi(B_i) \rck \\
  & = & \nu(B) \bE \lck \Phi(B_i) \lambda(x_o | \Phi_{S_{obs}}) \rck
\end{eqnarray*}
which leads to $\bE \lck \Phi(B_i) \lambda(x_o | \Phi_{S_{obs}}) \rck =
\frac{1}{\nu(B)} \bE \lck \Phi(B_o) \Phi(B_i) \rck$.
  \item whereas for $x_i = x_o$,
\begin{eqnarray*}
  \bE \lck \Phi^2(B_o) \rck &=& \bE \lck \bE \lck
  \Phi^2(B_o) | \Phi_{S_{obs}} \rck \rck = \bE \lck \Phi(B_o) \bE \lck \Phi(B_o) | \Phi_{S_{obs}}
   \rck \rck \\
  & =&  \bE \lck \Phi(B_o) \nu(B) \lambda(x_o | \Phi_{S_{obs}}) \rck \\
  & = & \nu(B) \bE \lck \Phi(B_o) \lambda(x_o | \Phi_{S_{obs}}) \rck
\end{eqnarray*}
which leads to $\bE \lck \Phi(B_o) \lambda(x_o | \Phi_{S_{obs}}) \rck =
\frac{1}{\nu(B)} \bE \lck \Phi^2(B_o) \rck$.
\end{itemize}

\noindent Thus, $\widetilde C_o = \frac{1}{\nu(B)} C_o$ (what is also obvious in
prediction) and we get
$$\mu = C^{-1} C_o
+ \dfrac{1- \Un^T C^{-1} C_o}{\Un^T C^{-1} \Un} C^{-1} \Un.$$
Interpolating $\Phi(B_o)/\nu(B)$ or
$\lambda(x_o | \Phi_{S_{obs}})$ leads to the same kriging weights.

\noindent
Finally,
$$\bE \lck \widehat \lambda(x_o | \Phi_{S_{obs}}) \rck  =  \bE \lck \sum_{x_i \in S_{obs}} \mu_i \dfrac{\Phi(B_i)}{\nu(B)} \rck
= \bE \lck \dfrac{\Phi(B_o)}{\nu(B)} \rck
 \xrightarrow[\nu(B) \to 0]{} \lambda(x_o | \Phi_{S_{obs}}) $$
shows that $\widehat \lambda(x_o | \Phi_{S_{obs}})$ is an asymptotically unbiased predictor of $\lambda(x_o | \Phi_{S_{obs}})$.
\finpreuve

\subsection{Properties of the interpolator}

In order to develop the variance of the kriging interpolator,
we use the following Neuman series (see e.g. \cite{petersen2012}) to inverse the covariance matrix $C$, which holds when $\nu(B)$ and $\lambda$ are small enough:
\begin{equation}\label{eq:invC}
 C^{-1}
 %= \dfrac{1}{\lambda \nu(B)} \left\lbrack \Id + \lambda \nu(B)  \sum_{k=1}^\infty (-1)^k \lambda^{k-1} H^k \right\rbrack
= \dfrac{1}{\lambda \nu(B)} \left\lbrack \Id
+ \lambda \nu(B) J_\lambda \right\rbrack,
\end{equation}
where a generic element of the matrix $J_\lambda$ is given by
\begin{multline*}
J_\lambda[i,j] = \sum_{k=1}^\infty (-1)^k \lambda^{k-1} %H^k$ and $H^k =
 \lp g(x_i,x_{l_1}) - 1 \rp \lp g(x_{l_{k-1}},x_j) - 1 \rp
 \\
 \times \int_{S_{obs}^{k-1}}  \prod_{m=1}^{k-2} (g(x_{l_m},x_{l_{m+1}}) -1)
\ \dd x_{l_1} \dots \ \dd x_{l_{k-1}}.
\end{multline*}

\begin{proposition}\label{prop:prop}
In estimation the variance of $\widehat \lambda(x_o | \Phi_{S_{obs}})$ is
\begin{multline}
\label{eq:varinterp}
  \var \lp \widehat \lambda(x_o | \Phi_{S_{obs}}) \rp =
  \dfrac{\lambda}{\nu(B)} + 2 \lambda^2  \un_{x_o}^T J_\lambda  \un_{x_o}
  + 2 \lambda^3 \nu(B)  \un_{x_o}^T J_\lambda (G_o - 1) \\
   +
   \lambda^3 \nu^2(B) (G_o - 1)^T(G_o - 1) + \lambda^4 \nu^3(B) (G_o - 1)^T J_\lambda (G_o-1) \\
    +
  \frac{1-\Big[1+ \lambda \nu(B)  \Un^T J_\lambda  \un_{x_o} + \lambda \nu(B) \Un^T (G_o - 1) + \lambda^2 \nu^2(B)
  \Un^T J_\lambda (G_o-1)\Big]^2}{\frac{\nu(S_{obs})}{\lambda}
  + \nu^2(B) \Un^T J_\lambda \Un}.
\end{multline}
which leads to the following approximation when $\nu(B)$ and $\lambda$ are small enough,
  \begin{equation}\label{eq:varest}
   \var \lp \widehat \lambda(x_o | \Phi_{S_{obs}}) \rp \approx \dfrac{\lambda}{\nu(B)}.
  \end{equation}
In prediction the variance reduces to
\begin{eqnarray}\label{eq:varest2}
   \var \lp \widehat \lambda(x_o | \Phi_{S_{obs}}) \rp & = &
   \lambda^3 \nu^2(B) (G_o - 1)^T(G_o - 1) \nonumber \\
& & + \lambda^4 \nu^3(B) (G_o - 1)^T J_\lambda (G_o-1) \\
 %  \lambda^3 \int_S (g(x_o,y)-1)^2 \dd y  \\
   & & +
  % \lambda^4 \int_{S^2} (g(x_o,y)-1) \widetilde J_\lambda(y,z) (g(x_o,z)-1) \dd y \dd z, \nonumber
  \dfrac{1-\Big[ \lambda \nu(B) \Un^T (G_o - 1) + \lambda^2 \nu^2(B)
  \Un^T J_\lambda (G_o-1)\Big]^2}{\frac{\nu(S_{obs})}{\lambda}
  + \nu^2(B) \Un^T J_\lambda \Un}. \nonumber
  \end{eqnarray}
\end{proposition}

\noindent
\textbf{Proof:} % of Proposition~\ref{prop:prop}:}
The variance of $\widehat \lambda(x_o | \Phi_{S_{obs}})$ is given by
\begin{eqnarray*}
  \var \lp \widehat \lambda(x_o | \Phi_{S_{obs}}) \rp &=&
  \var \lp \sum_{x_i \in S_{obs}} \mu_i \frac{\Phi(B_i)}{\nu(B)} \rp
  = \frac{1}{\nu^2(B)} \mu^T C \mu  \\
  & =& \frac{1}{\nu^2(B)} \lce C_o^T C^{-1} C_o + \dfrac{1 -
  (\Un^T C^{-1} C_o)^2}{\Un^T C^{-1} \Un} \rce.
\end{eqnarray*}

\medskip

\noindent
$\bullet$ When estimating the local intensity, i.e. for $x_o$ lying in the observation window, we have
$$C_o = \lambda \nu(B) \un_{x_o} + \lambda^2 \nu^2(B) (G_o-1).$$
Thus, from Equation~(\ref{eq:invC}):
\begin{eqnarray*}
  C_o^T C^{-1} C_o & = & \lambda \nu(B) \Big[ 1 + \lambda \nu(B) \lp J_\lambda(x_o,x_o)
  + 2 \un_{x_o}^T (G_o - 1) \rp  \\
  & & \lambda^2 \nu^2(B) \lp 2 \un_{x_o}^T J_\lambda (G_o - 1) +
  (G_o-1)^T (G_o-1) \rp \\
  & &  \lambda^3 \nu^3(B) (G_o -1)^T J_\lambda (G_o-1) \Big],
\end{eqnarray*}
where $J_\lambda(y,z) = \sum_{k=1}^\infty (-1)^k \lambda^{k-1} \int_{S_{obs}^{k-1}} (g(y,x_{l_1})-1)
 \prod_{m=1}^{k-2} (g(x_{l_m},x_{l_{m+1}}) -1) (g(x_{l_{k-1}},z)-1) \dd x_{l_1} \dots
 \dd x_{l_{k-1}}$,
\begin{eqnarray*}
  \Un^T C^{-1} C_o & = & 1 + \lambda \nu(B) \Big[ \Un^T J_\lambda \un_{x_o} + \Un^T (G_o-1)
  + \lambda \nu(B) \Un^T J_\lambda (G_o-1) \Big],
\end{eqnarray*}
and
\begin{eqnarray*}
  \Un^T C^{-1} \Un & = & \dfrac{1}{\lambda \nu(B)} \Big[ n_{obs}
  + \lambda \nu(B) \Un^T J_\lambda \Un \Big] = \dfrac{\nu(S_{obs})}{\lambda \nu^2(B)}
  + \Un^T J_\lambda \Un.
\end{eqnarray*}
Then, if $\nu(B)$ is very small,
$\dfrac{C_o^T C^{-1} C_o}{\nu^2(B)}$ varies in $\dfrac{\lambda}{\nu(B)}$ and
$\dfrac{1 - (\Un^T C^{-1} C_o)^2}{\nu^2(B) \Un^T C^{-1} \Un}$ in $\dfrac{\lambda}{\nu(S_{obs})}$. Thus, we get Equation~(\ref{eq:varest}).

\bigskip

\noindent
$\bullet$ When predicting the local intensity, i.e. for $x_o$ outside the observation window, we have
$C_o = \lambda^2 \nu^2(B) (G_o-1).$
Thus, from
$$  C_o^T C^{-1} C_o = \lambda^3 \nu^3(B) (G_o - 1)^T(G_o - 1)
+ \lambda^4 \nu^4(B) (G_o - 1)^T J_\lambda (G_o-1)$$
and
$$ \Un^T C^{-1} C_o = \lambda \nu(B) \Un^T (G_o - 1) + \lambda^2 \nu^2(B) \Un^T J_\lambda (G_o-1)$$
we get Equation~(\ref{eq:varest2}).

\finpreuve

\subsection{Defining an optimal mesh size}

When estimating the local intensity, the Integrated Mean Squared Error of $ \widehat \lambda(x | \Phi_{S_{obs}})$ leads to the following approximation~:
 \begin{eqnarray}
 \label{eq:IMSE}
 IMSE \lp \widehat \lambda(x | \Phi_{S_{obs}}) \rp & = & \int_{S} \lck \lp \lambda(x | \Phi_{S_{obs}}) -
 \bE[\widehat \lambda(x | \Phi_{S_{obs}} )] \rp^2 + \var \lp \widehat \lambda(x | \Phi_{S_{obs}} ) \rp \rck \dd x \nonumber \\
 & \approx & \dfrac{\sqrt{\nu(B)}}{12} \int_{S} \| \nabla \lambda(x | \Phi_{S_{obs}}) \|^2 \dd x
 + \dfrac{\lambda \nu(S)}{\nu(B)}.
 \end{eqnarray}
% This approximation can easily be obtained by considering the Taylor expansion of $\lambda(x | \Phi_{S_{obs}} )$.
\noindent
Then, we propose to find the optimal mesh of the interpolation grid
by minimising $IMSE \lp \widehat \lambda(x | \Phi_{S_{obs}}) \rp$ (see \ref{app:C}), and we get~:
\begin{equation}
\label{eq:optimalmesh}
\nu_{opt}(B) = \sqrt{\dfrac{12 \lambda \nu(S)}{\int_{S} \| \nabla \lambda(x | \Phi_{S_{obs}}) \|^2 \dd x}}.
\end{equation}

Note that because the optimal mesh depends on the inverse of squared $L_2$-norm of the gradient of the local intensity,
it decreases for clustered point patterns. Conversely, it increases for regular
point patterns.

In practice the optimal mesh can be approximated by estimating the gradient of the intensity over a fine grid (see \ref{app:simuls}).

\medskip

\noindent When predicting the local intensity, the smaller the mesh, the better.
Computation time is the only limit.

\section{Real case study}
\label{sec:real}

\begin{figure}[h]
\centering
\includegraphics[width=.8\linewidth]{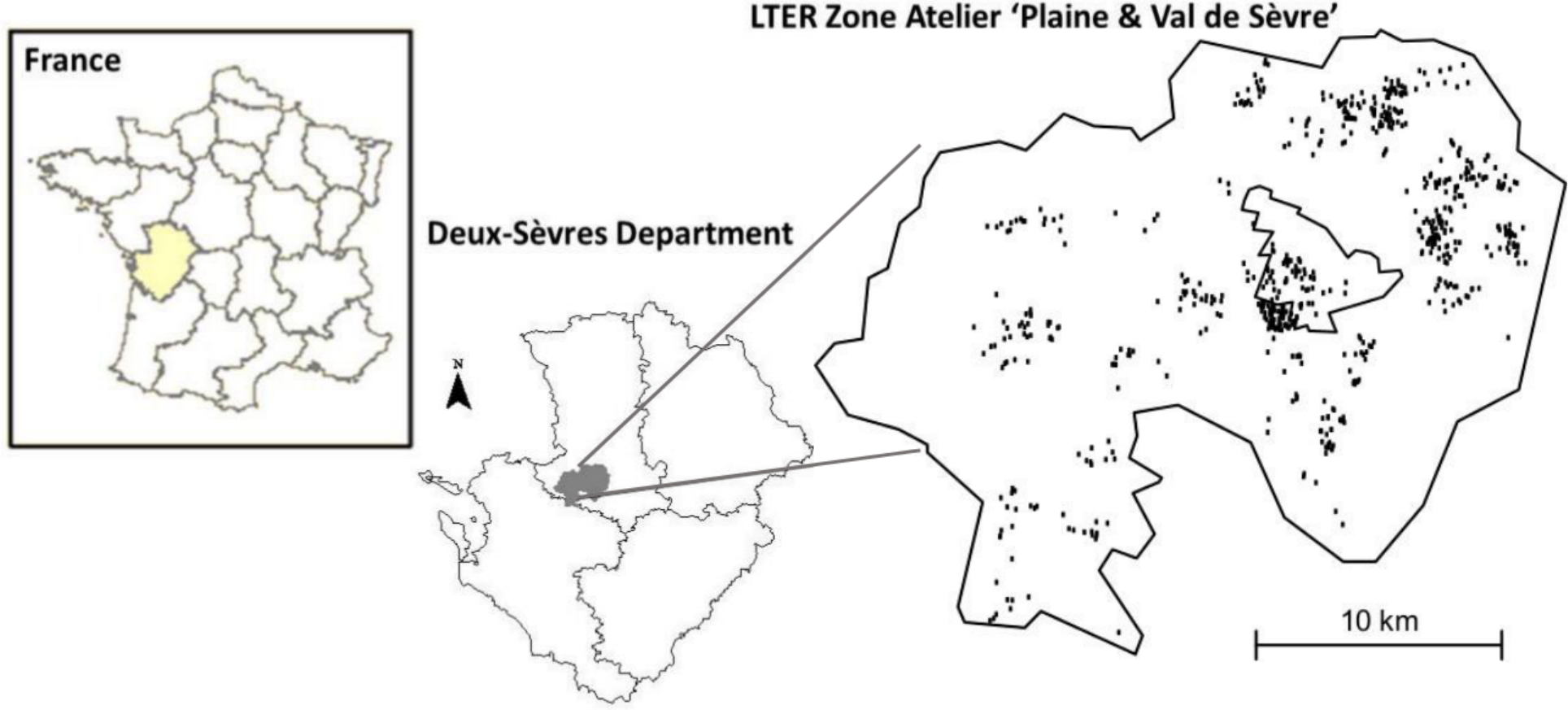}
\caption{Montagu's Harriers nest locations in the Zone Atelier "Plaine \& Val de S{\`e}vre".}
\label{fig:nests}
\end{figure}

In this section we estimate and predict the intensity of Montagu's Harriers' nest locations in the
Zone Atelier "Plaine \& Val de S{\`e}vre"\footnote{{\tt http://www.za.plainevalsevre.cnrs.fr/}} (Figure \ref{fig:nests}), a NATURA$2000$ site in France of $450$ km$^2$, designated for
its remarkable diversity of bird species.
Dots in Figure \ref{fig:nests} represent the exhaustive collection of Montagu's Harriers' nest locations. The area in the center of the Zone Atelier delineates the administrative boundaries of the commune Saint-Martin-de-Bernegoue, which will be used for prediction.

\subsection{Estimation of the pair correlation function}
\label{sec:pcfpractice}

The pair correlation is estimated as defined in \cite{stoyan1994}~:
$$ \widehat{g}(r) = \dfrac{1}{2 \pi r} \sum_{\xi \in \Phi_{S_{obs}}} \sum_{\zeta \in \Phi_{S_{obs}}}^{\ne} \dfrac{{\bf k}_h \left( r - \| \xi - \zeta \| \right)}{\text{prop} \left( S_{obs} \cap S_{obs,\xi-\zeta} \right)}$$
where ${\bf k}_h$ is the Epanechnikov kernel with bandwidth $h$, the optimal Stoyan's bandwidth equals to $0.15/\sqrt{\Phi(S_{obs})/\nu(S_{obs})}$ and
$\text{prop} \left( S_{obs} \cap S_{obs,\xi-\zeta} \right)$ is the proportion of translations of $(\xi,\zeta)$ which have both $\xi$ and $\zeta$ inside $S_{obs}$.
Figure~\ref{fig:estimnests}.a) shows the pair correlation function estimated from either all data point locations (solid line) or only the ones outside the boundaries of Saint-Martin-de-Bernegoue (dashed line). These estimates are characteristic of a Thomas cluster process with an infinite range of correlation, see \cite{illian2008}.

\subsection{Intensity estimation}
\label{sec:nuoptpractice}

For our kriging estimator, the optimal mesh is obtained by minimising the IMSE. Usual nonparametric estimation methods also require to preliminary set the smoothing parameter and this parameter is chosen as an optimal value minimising a specific criterion (mean square error,
integrated bias, asymptotic mean square error).
In our case, we have an explicit formula of the optimal mesh (Equation~(\ref{eq:optimalmesh})), which depends on the unknown terms $\lambda$ and $\lambda(x|\Phi_{S_{obs}})$.
If $\hat{\lambda}=\Phi(S_{obs})/\nu(S_{obs})$ appears to be a natural candidate to estimate $\lambda$, the challenging goal is to estimate
$\int_{S} \|\nabla \lambda(x|\Phi_{S_{obs}})\|^2 \dd x$.
Based on simulation experiments (\ref{app:simuls}), we consider a Gaussian kernel~\cite{silverman1986}, with a bandwidth minimising the mean-square error criterion defined by \cite{diggle1985}, to get a good approximation of the gradient of $\lambda(x|\Phi_{S_{obs}})$ on a $200\times200$ grid.
 This methodology applied to the real dataset leads to a value of $\nu_{opt}(B)$ equals to 23.19 hectares, % $231910.8$ square meters,
 which corresponds to a grid of $64 \times 53$ cells.

\begin{figure}[h]
\centering
{\scriptsize a) \hspace{3cm} \hspace{3.2cm} b)}

\includegraphics[width=.45\linewidth]{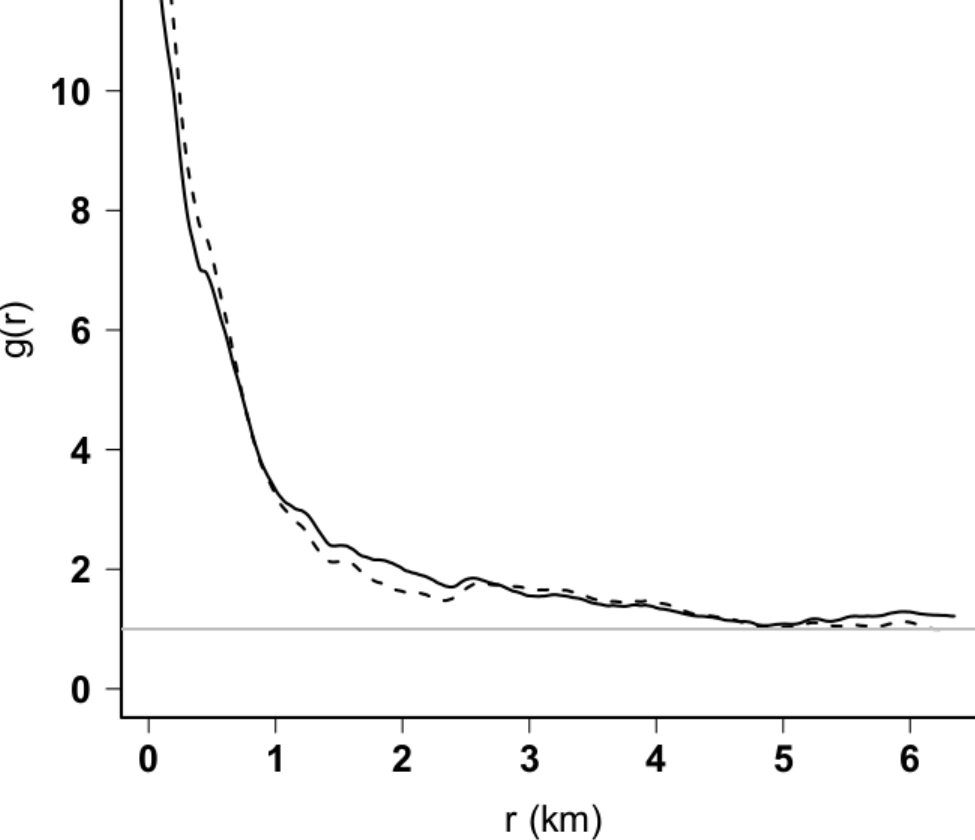}
\includegraphics[width=.45\linewidth]{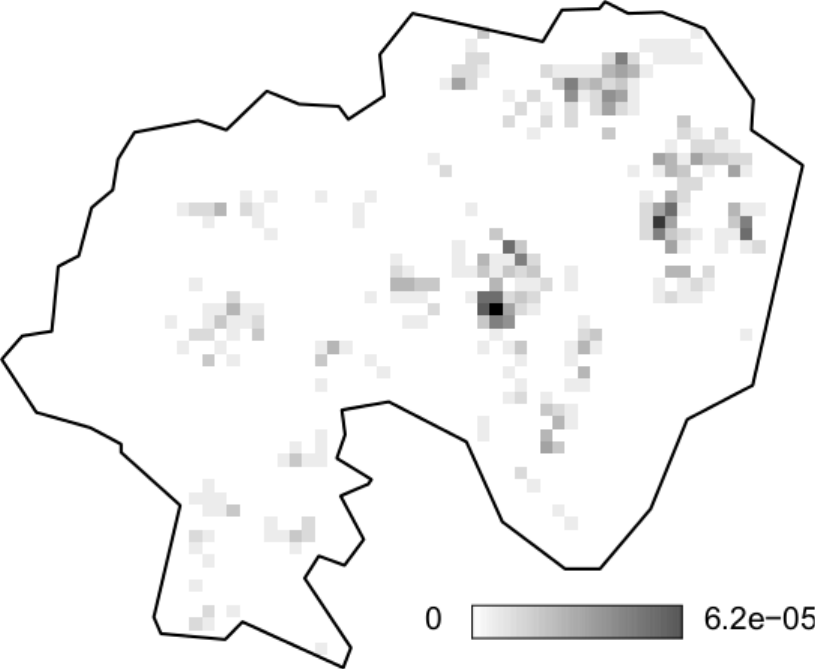}

{\scriptsize c) \hspace{3cm} \hspace{3.2cm} d)}

\includegraphics[width=.45\linewidth]{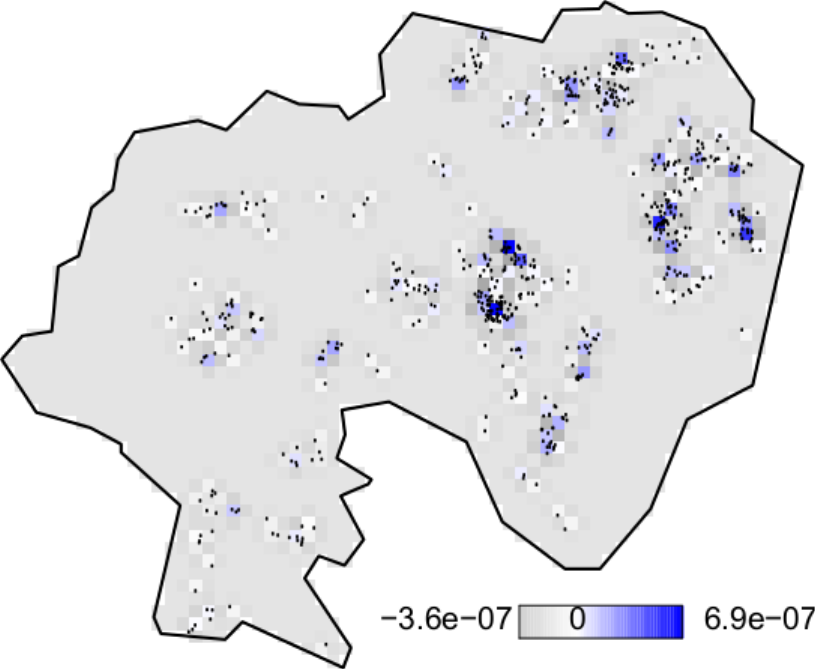}~
\includegraphics[width=.45\linewidth]{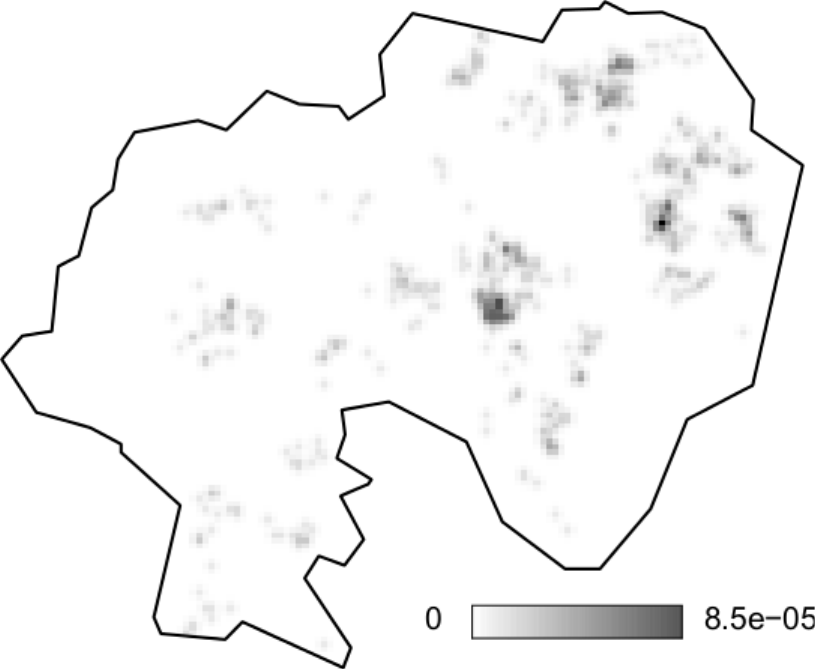}

\caption{a) Estimation of the pair correlation function from all points (solid line) and without those lying in Saint-Martin-de Bernegoue (dashed line).
Estimation of Montagu's Harriers nest locations using our kriging interpolator on the optimal grid (b) and using a Gaussian kernel on a $128 \times 128$ grid (d).
c) Difference between our estimator and the Gaussian kernel smoothing at the same resolution: dark grey (resp. blue) indicates higher values of the Gaussian kernel (resp. our kriging) estimates.}
\label{fig:estimnests}
\end{figure}

 Figure~\ref{fig:estimnests}.b) shows the kriging estimate on the optimal grid. Figure~\ref{fig:estimnests}.d) represents an estimate obtained by a Gaussian kernel, with a bandwidth selected as previously mentioned, on a $128 \times 128$ grid (default of the {\tt spatstat} function '{\tt density.ppp}', \cite{baddeley2005}). Figure~\ref{fig:estimnests}.c) illustrates the difference between our estimation and the one obtained by Gaussian kernel smoothing at the same grid resolution.
  Our kriging interpolator gives higher values of the local intensity (in blue in Figure~\ref{fig:estimnests}.c)) close to aggregated observation points than the kernel estimator, while the maximum value may be higher for the later.
  This illustrates that our method may be particularly relevant for point patterns strongly aggregated at a small scale.

\subsection{Intensity prediction}

\begin{figure}[h]
\centering
\includegraphics[width=.6\linewidth]{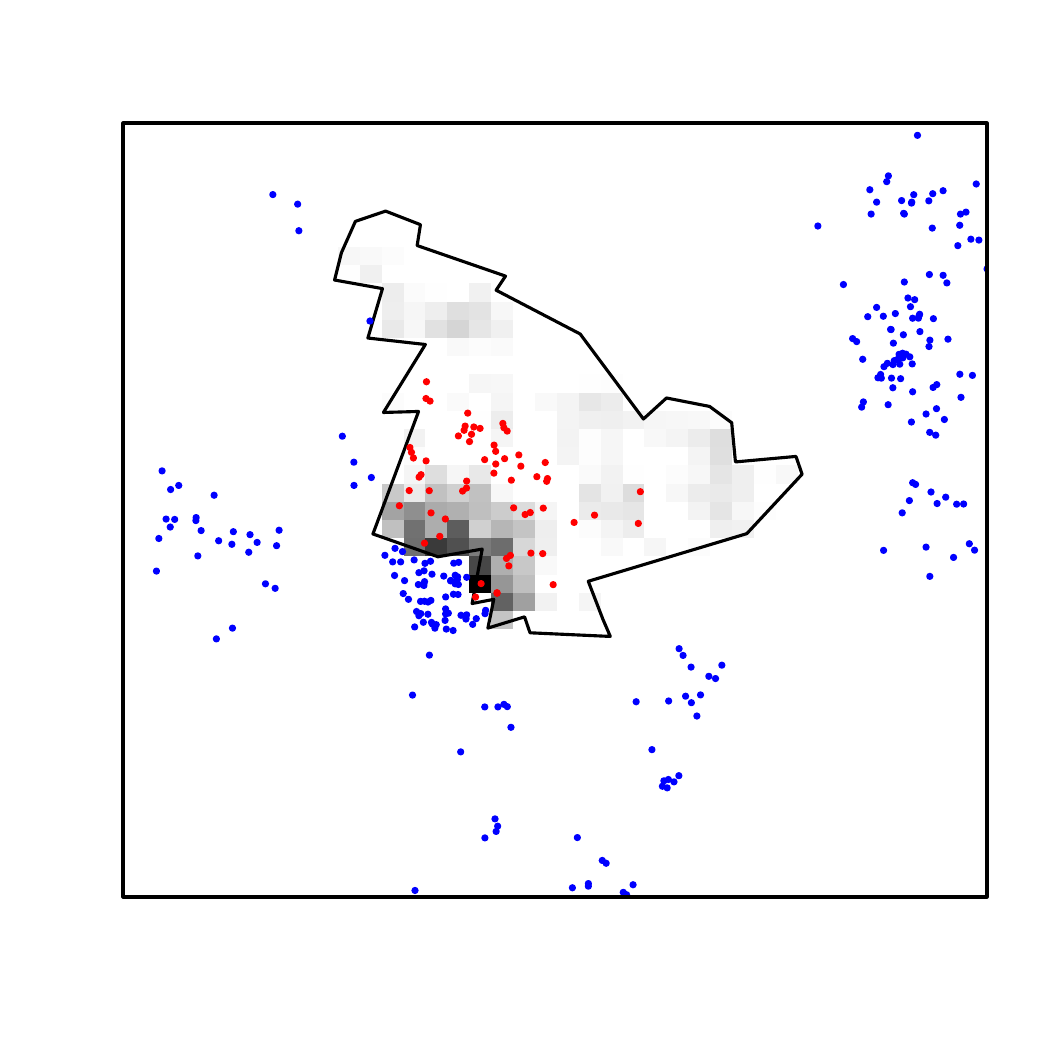}

\caption{Prediction within the commune of Saint-Martin-de-Bernegoue.}
\label{fig:predictnests}
\end{figure}

In order to apply our kriging predictor to the real dataset, we consider an unobserved window $S_{unobs}$ defined by the administrative boundaries of the commune Saint-Martin-de-Bernegoue in the center of the 'Zone Atelier' (Figure \ref{fig:nests}). Thus, we remove the points in this area (red dots in Figure \ref{fig:predictnests}) and use the remaining nest locations (blue dots in Figure \ref{fig:predictnests}) to predict the local intensity within $S_{unobs}$. We consider a grid of size $100 \times 100$ over $S$ to make the prediction. The estimated pair correlation function is plotted in Figure~ \ref{fig:estimnests}.a) (dashed line).
The result, zoomed in Figure~\ref{fig:predictnests}, shows that the kriging predictor is able to reproduce the second-order structure of the underlying point process. In particular, it reproduces clusters as soon as there are points close  enough to the boundary of the unobserved area.
 This will be further illustrated and discussed in the next section.
Note that at distances greater than the range of the pair correlation function, the method can only provide a constant intensity estimate.

\section{Illustrative simulation experiments}
\label{sec:experiments}

\subsection{Objectives}

Now, we focus on the kriging predictor and explore its accuracy through simulation
experiments, varying rate and shape of the observation window $S_{obs}$. To measure the quality
of prediction, we compute the mean bias (MB) and the mean square error of
prediction (MSEP):
 \begin{eqnarray*}
MB &=& \frac{1}{n_{obs}} \sum_{x \in \Phi_{S_{obs}}} \frac{1}{n_{sim}} \sum_{k=1}^{n_{sim}} \left( \widehat{\lambda}_k(x|\Phi_{S_{obs}}) - \lambda_k(x|\Phi_{S_{obs}}) \right), \\
MSEP &=& \frac{1}{n_{obs}} \sum_{x \in \Phi_{S_{obs}}} \frac{1}{n_{sim}} \sum_{k=1}^{n_{sim}} \left( \widehat{\lambda}_k(x|\Phi_{S_{obs}}) - \lambda_k(x|\Phi_{S_{obs}}) \right)^2,
\end{eqnarray*}
\noindent where $\lambda_k$ and $\widehat{\lambda}_k$ correspond respectively to the intensity and its predictor on the $k$th simulation and $n_{sim}$ is the number of simulations.
 We also compute the coefficient of determination $R^2$ of the regression between the predicted values of the local intensity and the theoretical ones.

\subsection{Experimental design}

Throughout our experimental study, in order to simplify the analysis of the two parameters
of interest (rate and shape of $S_{obs}$), we decide to simulate all
point patterns from a single spatial point process model. We consider in the sequel a Thomas
process, for which we have explicit formulas of the intensity, pair correlation function and others characteristics
(see \cite{illian2008}, p.377)~:
\begin{eqnarray}
\label{eq:lambdatheo}
\lambda(x|\Phi_{S_{obs}}) &= &\sum_{\xi \in \Phi_{S_{obs}}} \frac{\mu}{2\pi\sigma^2} \exp\left(-\frac{\|x-\xi\|^2}{2 \sigma^2}\right)\text{, for all }x \in S, \\
g(r) &=& 1 + \frac{1}{4\pi\kappa\sigma^2} \exp \left(- \frac{r^2}{4\sigma^2} \right)\text{, for } r \geq 0. \nonumber
\end{eqnarray}
 Such a Cox model is of interest as it
models spatial aggregation, a condition often observed in practical situations of intensity prediction.
We simulate $n_{sim}=1000$ patterns of a Thomas process in the unit square with parameters:
\begin{itemize}
\item $\kappa=10$, the intensity of parent points from a homogeneous Poisson point process,
\item $\mu=50$, the mean number of children points around each parent point from a Poisson distribution,
\item $\sigma=0.05$, the standard deviation of the gaussian density distribution centered at each parent point.
\end{itemize}

\begin{figure}[h]
\centering
\includegraphics[width=.6\linewidth]{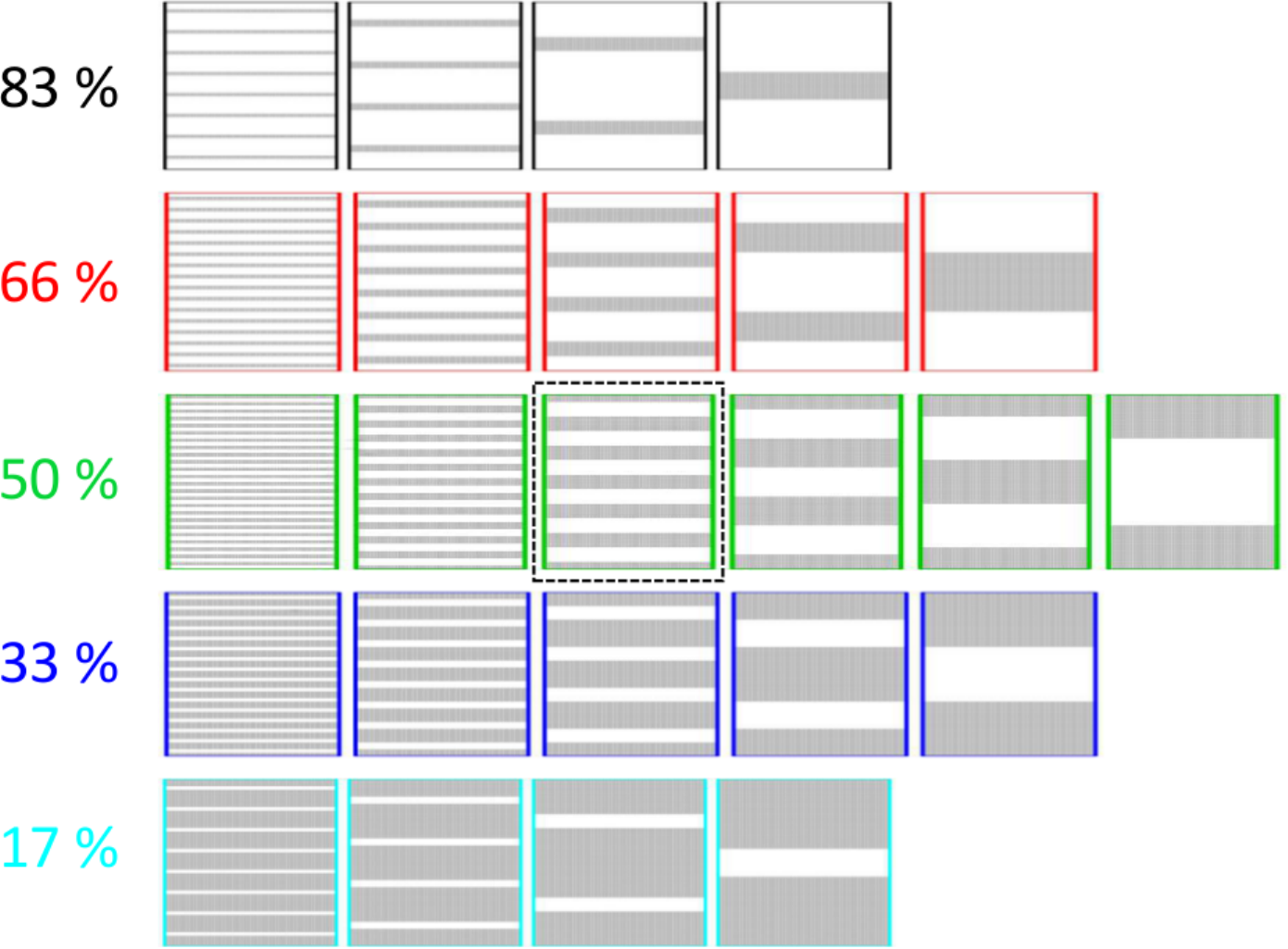}
\caption{Windows of interest with $S_{obs}$ (resp. $S_{unobs}$) the union of white (resp. grey) bands.}
\label{fig:sunobs}
\end{figure}

Several windows of interest $S_i$, with $i=1,\cdots,24$, are considered (Figure~\ref{fig:sunobs}),
corresponding to different observation rates (83\%, 66\%, 50\%, 33\% and 17\%).
The unobserved windows $S_{unobs}$ are defined by the union of bands, with varying width (in grey in Figure~\ref{fig:sunobs}).

Because the weights in our kriging interpolator depend on the pair correlation function, in our experiment we compare results arising from the theoretical pair correlation function, and from its estimate defined in Section~\ref{sec:pcfpractice}.
In order to estimate the pair correlation function from similar number of points in each window $S_i$,
we first simulated point patterns within a larger window (the initial one extended on the right side), so that the area of observation zones equals to one. The pair correlation function is then estimated from this first pattern and the prediction is made on its restriction to the initial unit square.

\subsection{Results}
\begin{figure}[h]
\centering
{\scriptsize MB \hspace{4.5cm} MSEP}

\includegraphics[width=.4\linewidth]{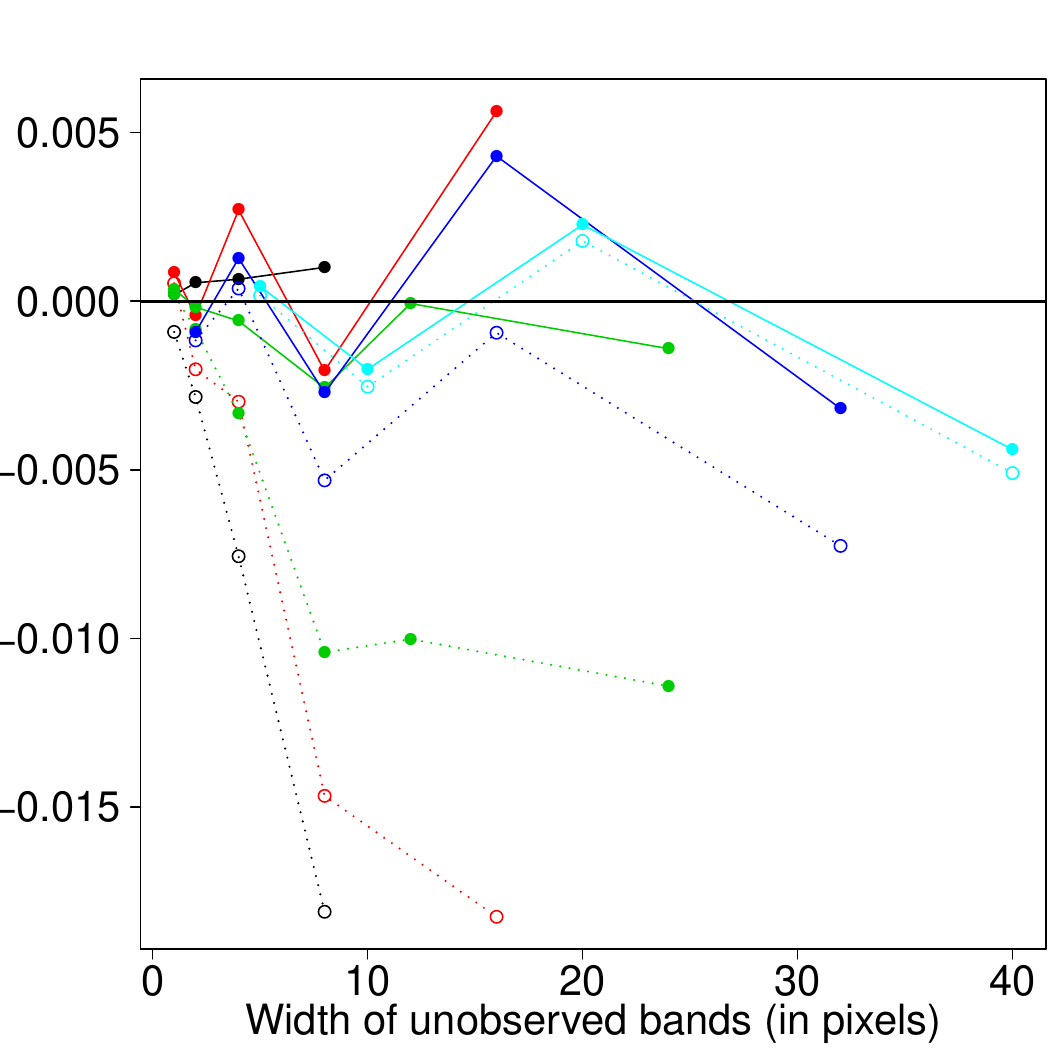}~
\includegraphics[width=.4\linewidth]{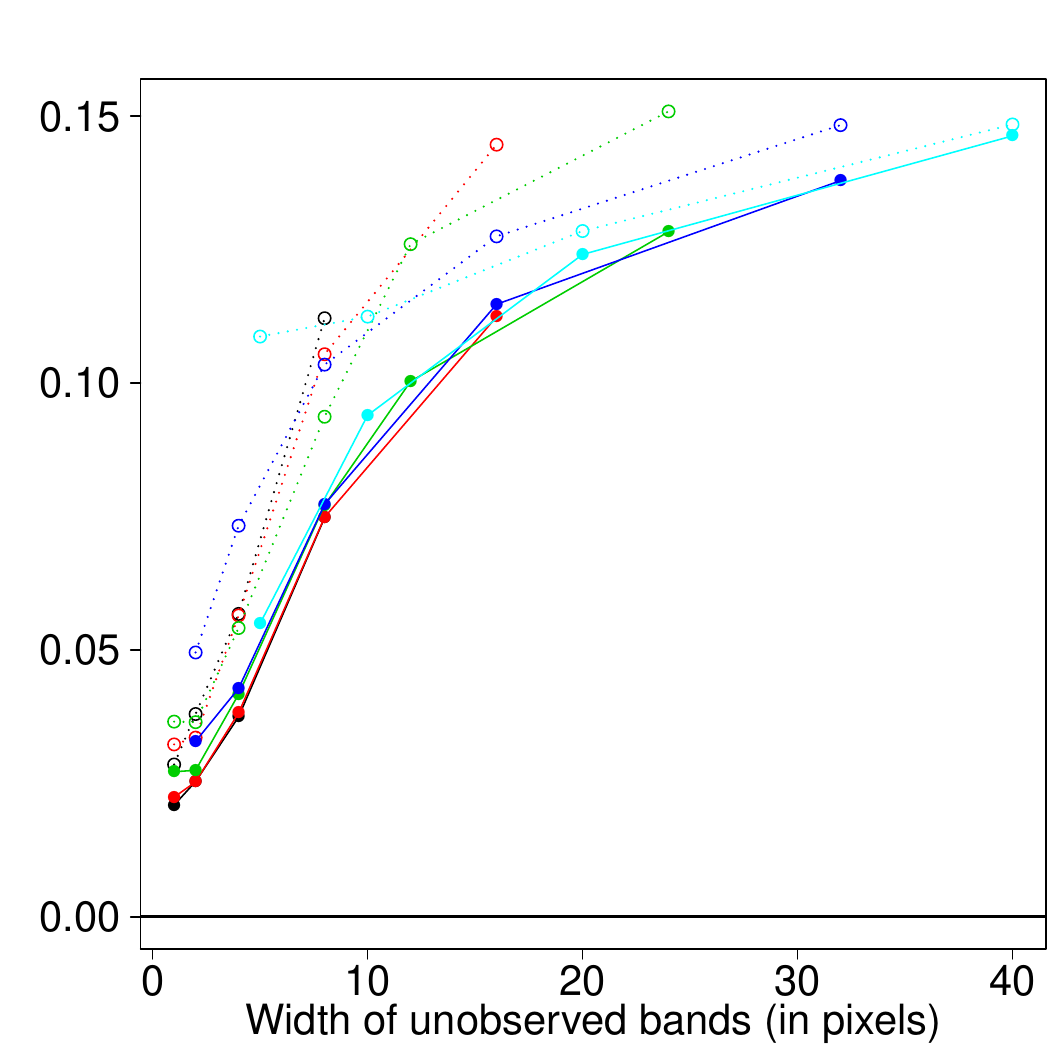}
\caption{Mean Bias (left) and Mean Square Error of Prediction (right) of our kriging predictor related to the width of the unobserved bands,
and to the observation rates~: 83\% in cyan, 66\% in blue, 50\% in green, 33\% in red and 17\% in black. The lines correspond to a linear approximation of the MSEP values when $g$ is known (solid lines) or estimated (dotted lines).}
\label{fig:MBMSEP}
\end{figure}

The mean bias and the mean square error of prediction are presented in Figure~\ref{fig:MBMSEP}, with theoretical values of the pair correlation function (solid lines) or an estimate (dotted line).
It shows that the mean bias has no effect on the MSEP.
With theoretical values of the pair correlation function,
the mean bias is close to zero whatever the width of the bands defining $S_{unobs}$, which numerically reveals the unbiasedness statistical property of our predictor. When the pair correlation function is estimated, $\lambda_k$ is under-estimated and the discrepancy is higher when the observation rate decreases than when the width of the unobserved bands increases.

At a given observation rate, the MSEP increases when the width of the unobserved bands increases. Indeed, the
geometry of our windows of interest implies that wider the unobserved bands, less numerous they are.
Consequently, for some simulated patterns, cluster points can completely fall within an unobserved
band, what damages the quality of prediction. At a given value of the unobserved band, we obviously see a slight
increase of the MSEP when the observation rate decreases.

\medskip

\begin{figure}[h]
\centering
{\scriptsize a) \hspace{3.3cm} b) \hspace{3.3cm} c)}

\includegraphics[width=.3\linewidth]{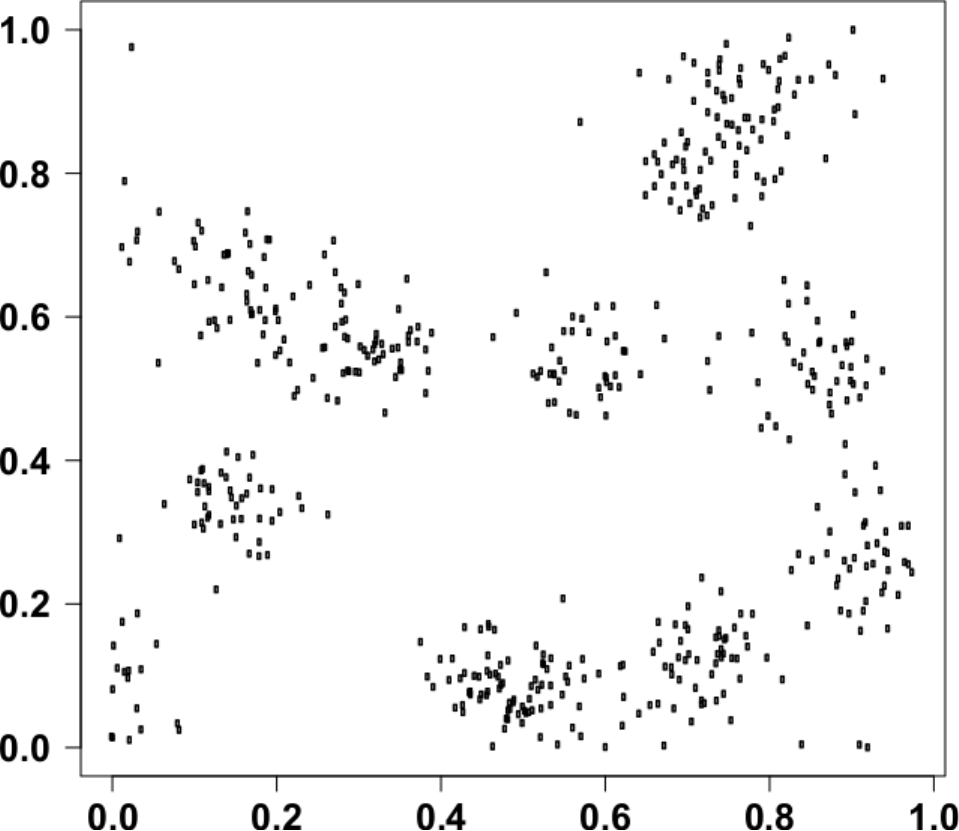}~\includegraphics[width=.3\linewidth]{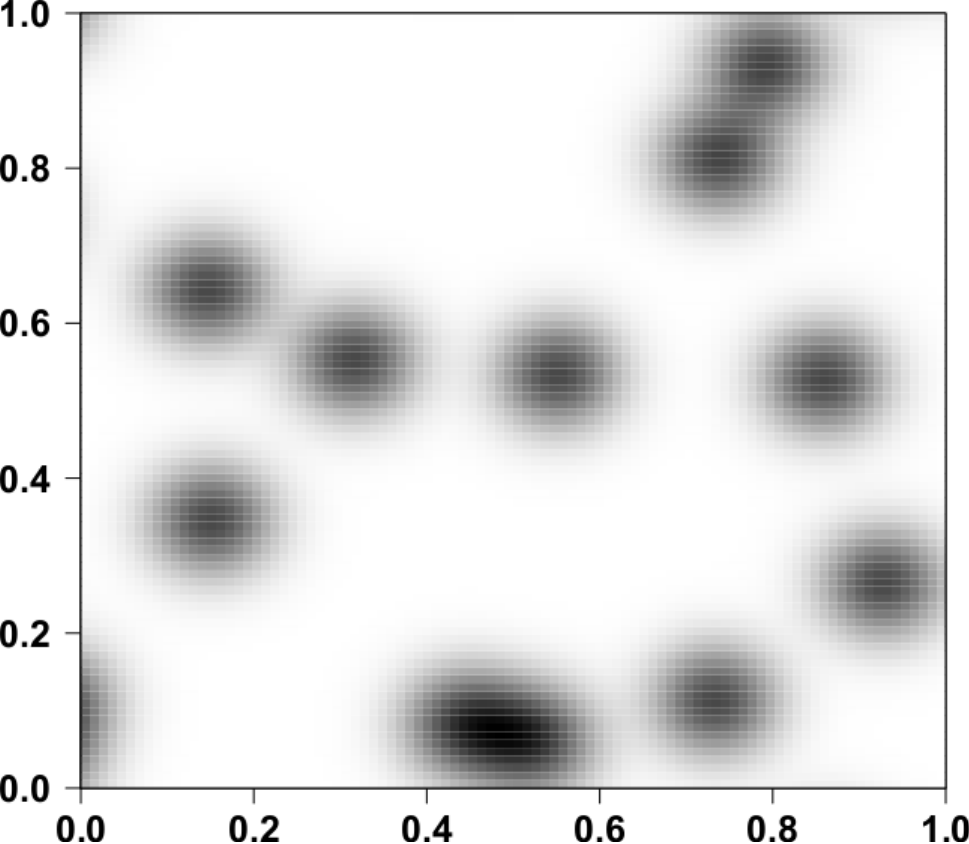}
~\includegraphics[width=.3\linewidth]{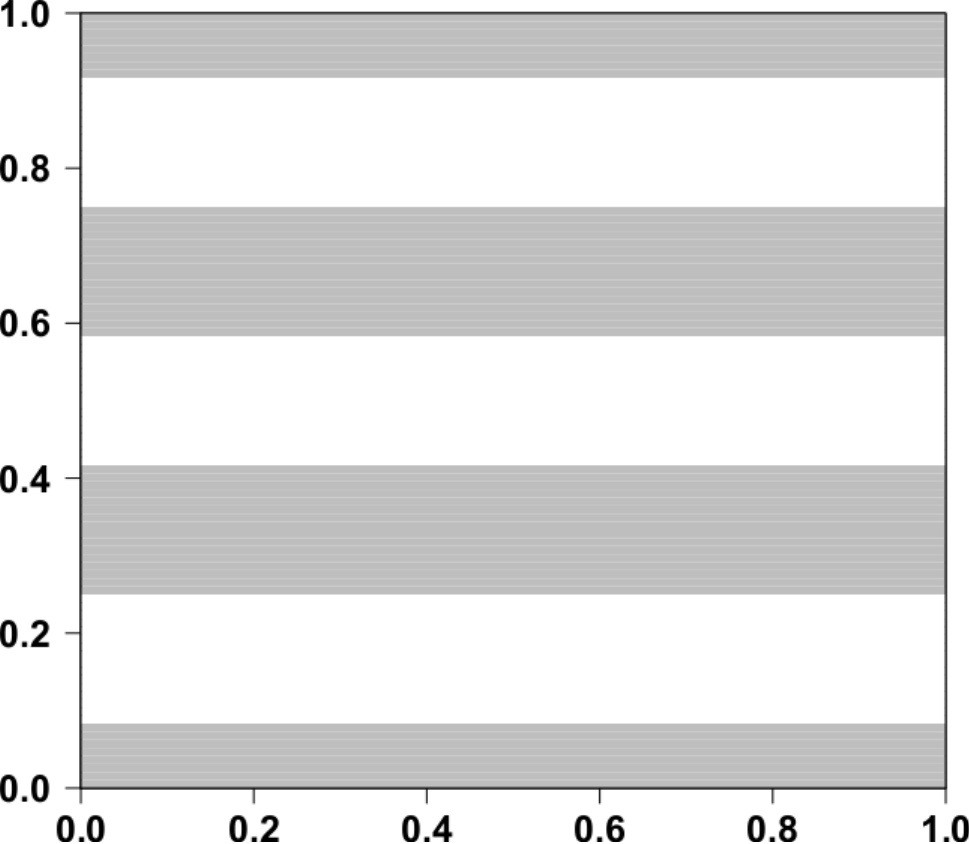}

{\scriptsize d) \hspace{3.3cm} e) \hspace{3.3cm} f)}

\includegraphics[width=.3\linewidth]{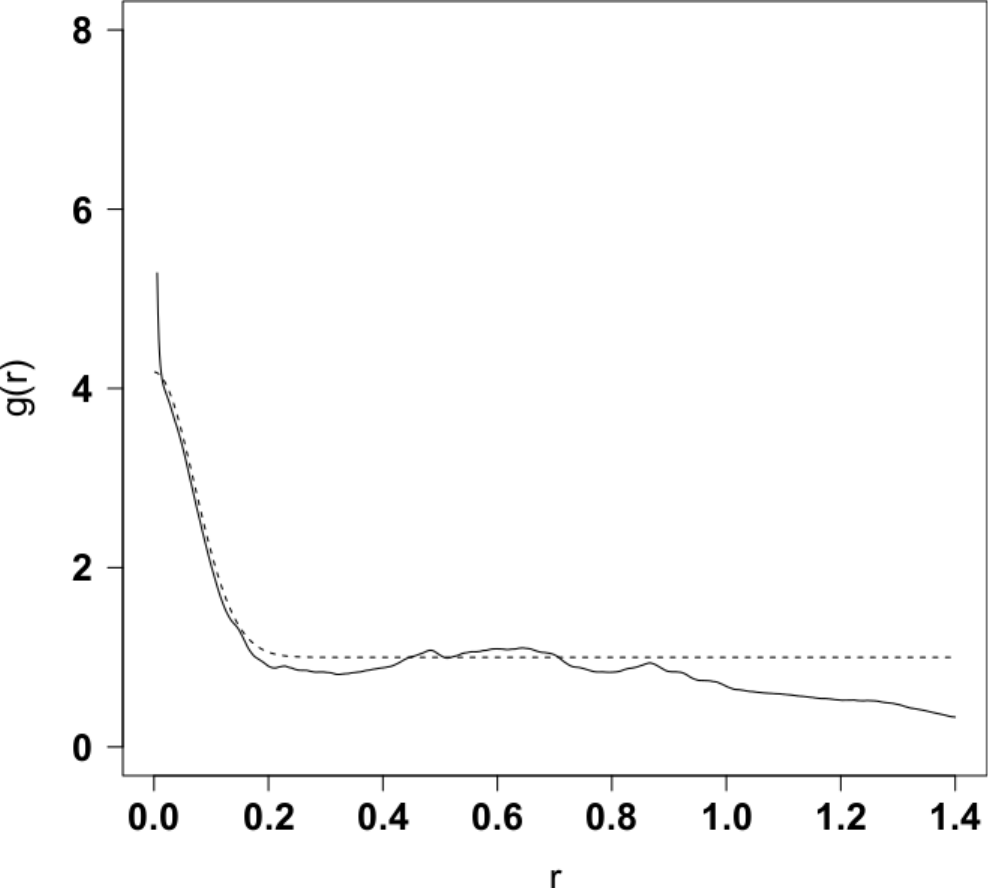}~\includegraphics[width=.3\linewidth]{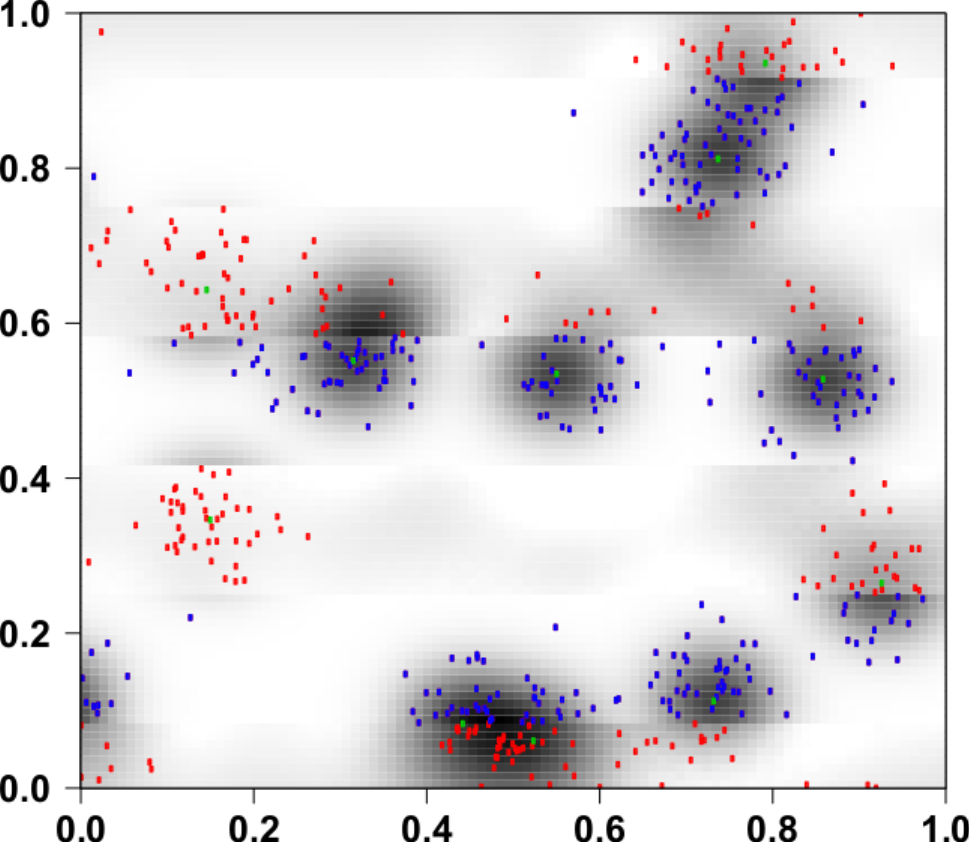}
~\includegraphics[width=.3\linewidth]{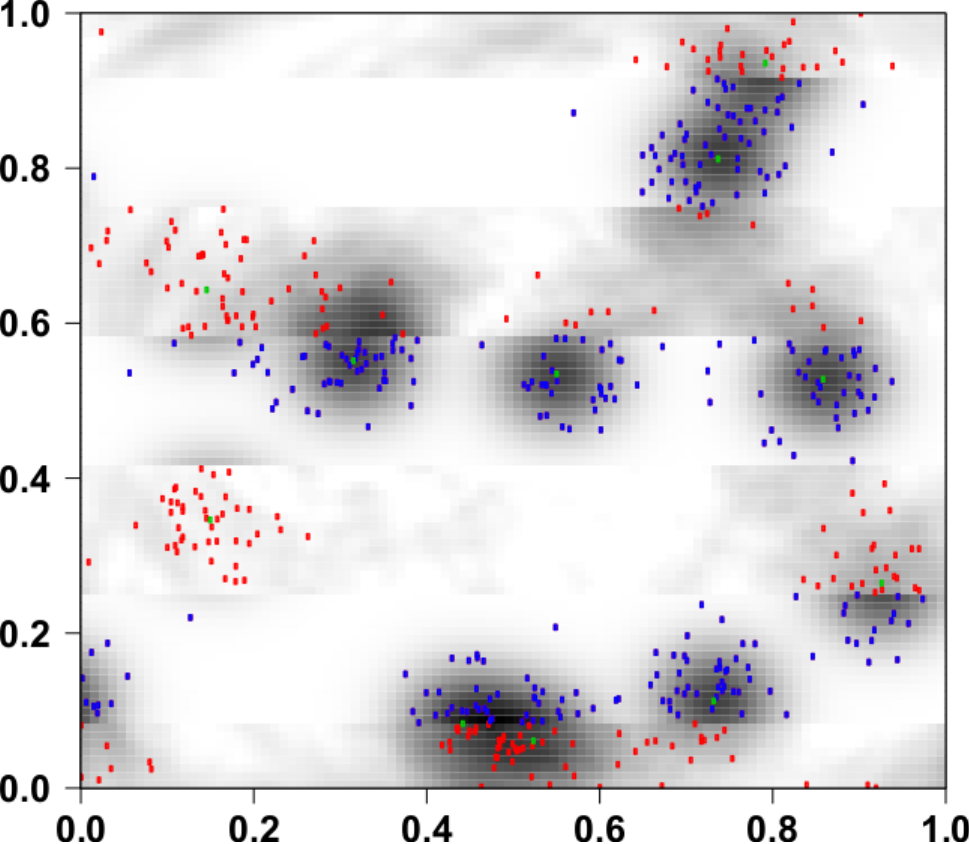}
\caption{a) Simulated pattern from a Thomas process (parent points in red and children points in black).
b) Theoretical intensity from the simulated pattern.
c) Observed window (light grey) and unobserved window (red) with an observation rate of $50\%$.
d) Theoretical (dotted line) and estimated (solid line) pair correlation function $g$.
e) and f) Theoretical intensity in the observed window and predicted local intensity
in the unobserved window obtained with the true pair correlation function (e) and an estimate (f).
}
\label{fig:example}
\end{figure}

We first illustrate the influence of the estimation of the pair correlation function onto the accuracy
of prediction on a single simulation. The simulated pattern, the associated theoretical intensity and the observation
window, with a rate of $50\%$ of observed areas, are represented in Figures \ref{fig:example}.a) to c) respectively.
The theoretical (dotted line) and estimated (solid line) pair correlation function are given in Figure~\ref{fig:example}.d).
Figures~\ref{fig:example}.e) and f) illustrate the theoretical local intensity in $S_{obs}$ and the prediction
in $S_{unobs}$ on a $96 \times 96$ grid, using the true (e) and the estimated (f) pair correlation function. In the first case, the prediction is relatively smooth and gives accurate results.
In the second case, the prediction is more noisy, but recover the same blocks with high intensity values.
In both cases, the method correctly predict the clusters when there are observations close to the unobserved bands. That is the case for all clusters located at the right hand side of the vertical line $x=0.25$. When the full cluster falls in the unobserved band, as the ones located at the left hand side of the vertical line $x=0.25$ the method fails in predicting the cluster.

\medskip

\begin{figure}[h]
\centering
{\scriptsize a) \hspace{4cm} b)}

\includegraphics[width=.35\linewidth]{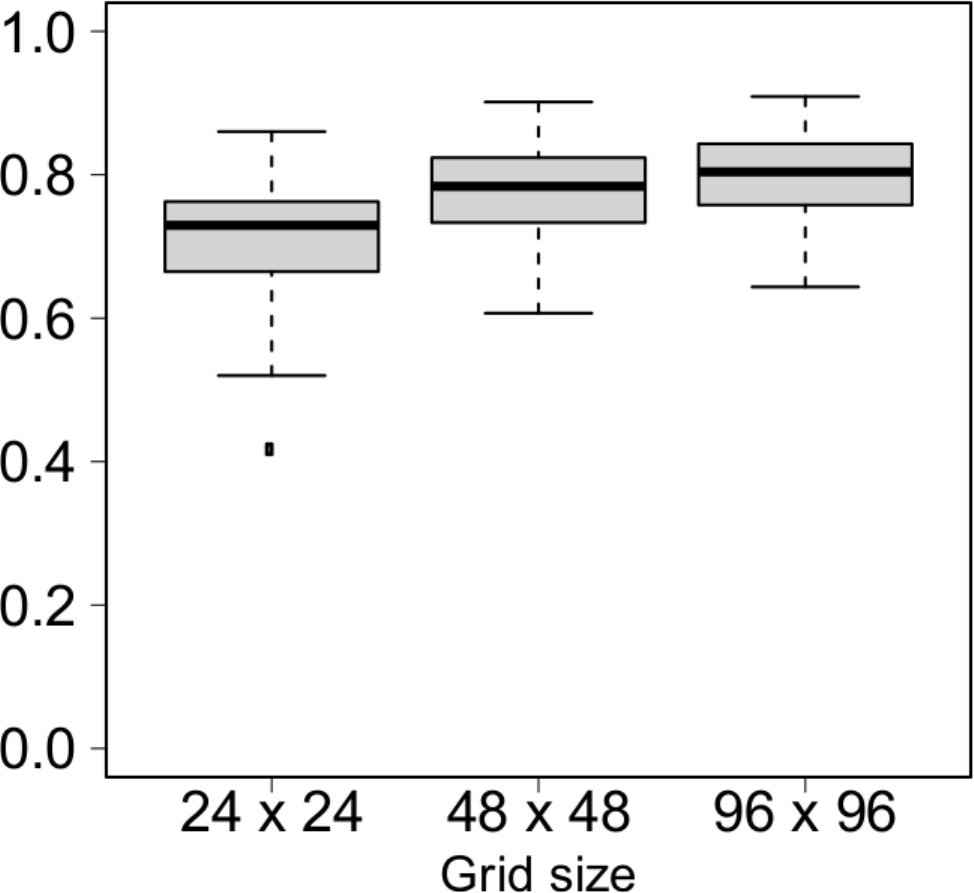}~\includegraphics[width=.35\linewidth]{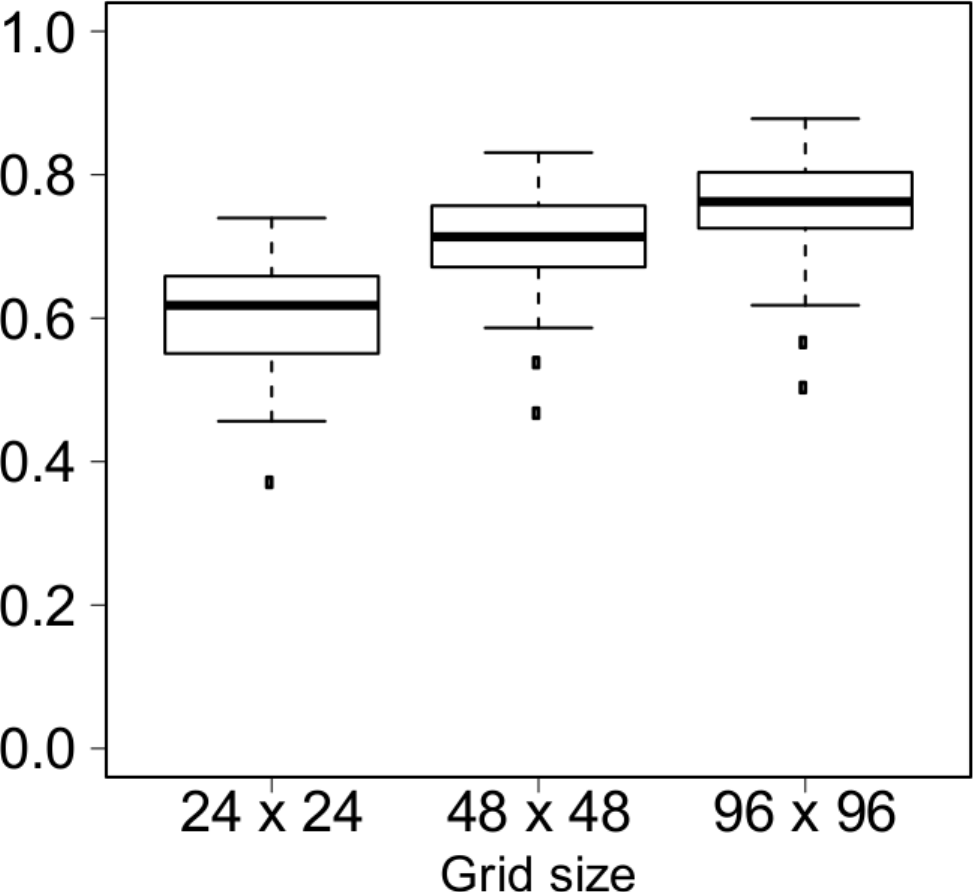}
\caption{$R^2$ in linear regressions of the predicted and theoretical values
of the local intensity, associated with different grid size, when the predictions are based on the theoretical
 pair correlation function (a) or an estimate (b).}
\label{fig:R2}
\end{figure}

\noindent
We plotted (Figure~\ref{fig:R2}) the boxplot of the coefficient of determination resulting from 100 simple linear regressions
between the predicted values of the local intensity and the theoretical ones, for different grid size ($24 \times 24$,
$48 \times 48$ and $96 \times 96$),
when the pair correlation function is estimated (Figure~\ref{fig:R2}.b)) or not (Figure~\ref{fig:R2}.a)). These results are related to an observation rate of 50\%, according to the window configuration highlighted Figure~\ref{fig:sunobs}.
We obviously see that the goodness of prediction increases when the grid resolution increases and when
the pair correlation function is known.
We considered a $96\times96$ grid as it is a trade off between computation times and a small mesh, allowing a good description of the intensity variations due to clusters. We obtained, in the
worst case where the pair correlation function is estimated, that the coefficient of determination $R^2$ is around $0.8$ (median).

\section{Discussion}

Our kriging method introduced to estimate and/or predict the local intensity of a stationary and isotropic
point process has a large number of advantages, particularly in prediction.
 Taking into account the spatial structure of the point pattern allows to perform the intensity estimation for point processes highly aggregated at fine scale. In the prediction framework, our kriging method is innovative
for the interpolation of the local intensity and presents good statistical properties (unbiasedness, low
variance...) when the pair correlation function is known. When it is estimated, the quality of our interpolator is
slightly reduced but our results can be improved by better taking into account
the double estimation of the pair correlation function and the local intensity on the same pattern.
%% comparaison avec les autres
This prediction method
is less time consuming than the reconstruction methods and appears a promising way in prediction of intensity
of a spatial point pattern.
Note that existing prediction methods are constrained within a class of point processes (Cox processes, \cite{monestiez2006,diggle2007,bellier2013,diggle2013}), making any comparison with our method very restrictive relatively to its broad scope of applications.
%% non sta
That is for instance the case of any point process obtained by a weak dependent process (e.g. Thomas, Markov) with a parameter driven by a stationary random field at a larger scale (e.g. Cox), but not only. Relaxing the stationary assumption implies to make further assumptions. For instance, if we consider a Cox process with intensity $\Lambda(x)$, the local intensity can be written as a sum,
$\lambda(x) + Y(x)$, or a product $\lambda(x) Y(x)$, where $Y(x)$ is a random field, centered or not, driven by $\Lambda(x)$ or not.
The formalism should be quite similar to the one of this paper, but with some confounding effects as for instance the ones observed
when using the same point pattern to estimate both a spatially varying intensity and second-order characteristics \cite{diggle2007b,gabriel2014}. One could thus allocate the effects at different scales.

\bigskip

%% Numerical guidelines

In our simulations and application, we used the {\tt R} function {\tt solve}, based on the LU factorization, to compute the inverse of the covariance matrix $C$.
This matrix is of dimension the square of the number of cells of the grid superimposed on the observation window. Thus, it can quickly become heavy to inverse. In such cases, estimating the matrix $C^{-1}$ using Equation~(\ref{eq:invC}) would be somewhat cumbersome. Thus, we propose instead to inverse the covariance matrix numerically.
Several approximations could be used, depending mainly on the width of $B$ with respect to $\lambda$
and the curvature of the pair correlation function:
 \begin{enumerate}
   \item if the diameter of $B$ is large, the covariance between two tiles at distance $r$ is equal to
            $\lambda \nu(B) \un_{r=o} +  \lambda^2 \int_{B \times D} \left( g(r+x-y) - 1 \right) \dd x \dd y$.
            It can be approximated numerically by computing for example the integral on a fine grid.
            Then the  finer the grid, the smaller the difference between the exact values and the  approximations,
            but the computing time cost can become prohibitive.
   \item when the diameter of $B$ becomes small, the integral can be approximated by
            $\nu(B)^2 g(r)$ so that the covariance is approximated by
            $\lambda \nu(B) \un_{r=o} + \nu(B)^2 \lambda^2 g(r)$,
   \item when the diameter becomes very small, $C$ may be approximated by $\lambda \nu(B) \Id$,
            a situation seldom met in practice, since it needs a tile $B$ small enough to neglect
            point dependence.
  \end{enumerate}
Approximation 2) will thus be the most reasonable one, needing only a $B$  small enough to consider
that $g(u+x)$ is almost constant for $x \in B$, but avoiding too small $B$ leading to large matrix inversion time.

Our estimation is roughly pixellated compared to kernel methods, but it does not oversmooth the intensity of highly aggregated point processes. We could take the benefit of the two approaches to get smoother estimations. Our on-going work consists in regularizing the counting process by a kernel and in defining a kriging estimator for the related random field.
Our optimal grid could then be used to define an optimal bandwidth, thus eliminating the Poisson aspect of classical kernels.

%% Covariables

\bigskip

Our method provides good predictions in areas at small distances of data locations. From the definition of the kriging predictor, at distances larger than the range of interaction, it only provide a constant mean value. To improve it and make it more relevant in practice, we could consider further information provided by covariates. From our application point of view, wheat field mapping could be of interest as Montagu's Harriers nest in there. From a methodological point of view, including covariates would imply that we should either consider external drift kriging (or any other universal kriging) rather than ordinary kriging); or spatial regression.

Finally, our kriging predictor depends on the count data in the grid cells, $B_i$,
and not on exact data locations in $B_i$. Thus we can further consider count data sets, as it is often the case in biodiversity measures, e.g. plant species abundance. The exact position of each plant is rarely given, but we know its abundance per small unit areas. So, once the pair correlation function is estimated from the point data subset, one can apply our method to interpolate the intensity.

% \section*{acknowledgments}

\section*{References}

\bibliography{mybibfile}

\appendix

\section{Proof of Lemma~\ref{lem:pp}}
\label{app:A}
\noindent
\begin{align*}
1) \ \ \bE \lck \Phi(B) \Phi(D) \rck & =  \bE \lck \lp \sum_{x \in \Phi_S} \un_{B}(x) \rp
  \lp \sum_{y \in \Phi_S} \un_{D}(y) \rp \rck \\
  &= \bE \lck {\sum\sum}_{x,y \in \Phi_S} \un_{B}(x) \un_{D}(y) \rck \hspace{2cm} \phantom{.} \\
  & =  \bE \lck \sum_{x \in \Phi_S} \un_{B}(x) \un_{D}(x) \rck +
  \bE \lck {\sum\sum}_{x \neq y \in \Phi_S} \un_{B}(x) \un_{D}(y) \rck  \\
  & =  \bE \lck \sum_{x \in \Phi_S} \un_{B \cap D}(x) \rck +
   \int_{B \times D} \lambda_2(x,y) \dd x \dd y \\
  & =  \int_{B \cap D} \lambda(x) \dd x + \lambda^2 \int_{B \times D} g(x,y) \dd x \dd y \\
  & =  \lambda \nu(B \cap D) + \lambda^2 \int_{B \times D} g(x-y) \dd x \dd y
\end{align*}
The following convergence result
\noindent
$\bP \lck \lce \Phi(B) = 1 \rce \cap \lce \Phi (D) = 1 \rce \rck
 = \lim_{\nu(B),\nu(D) \to 0} \bE \lck \Phi(B) \Phi(D) \rck$ ends the proof.

\medskip

\noindent 2) The proof idea is identical to 1).

\noindent
\begin{align*}
3) \ \  \var \lck \Phi(B) \rck & = \bE \lck \Phi^2(B) \rck - \bE^2 \lck \Phi(B) \rck
= \lambda \nu(B) + \lambda^2 \int_{B \times B} g(x-y) \dd x \dd y - \lp \lambda \nu(B) \rp^2 \hspace{1.5cm} \\
&= \lambda \nu(B) + \lambda^2 \lp \int_{B \times B} g(x-y) \dd x \dd y - \nu^2(B) \rp \\
&= \lambda \nu(B) + \lambda^2 \int_{B \times B} \lp g(x-y) -1 \rp \dd x \dd y
\end{align*}

\noindent
4) Let $B$ and $D$ so that $B \cap D = \emptyset$,
\begin{align*}
\cov \lp \Phi(B), \Phi(D) \rp & = \bE \lck \Phi(B) \Phi(D) \rck - \bE \lck \Phi(B) \rck \bE \lck \Phi(D) \rck\\
&= 0 + \lambda^2 \int_{B \times D} g(x-y) \dd x \dd y - \lambda^2 \nu(B)\nu(D) \\
&=\lambda^2 \int_{B \times D} \lp g(x-y) -1 \rp \dd x \dd y
\end{align*}
\finpreuve

%\newpage
\section{Proof of Proposition~\ref{prop:ppvsgeostat}}
\label{app:B}

\noindent 1) $m = \bE \lck Z(x) \rck = \bE \lck \Phi(B) \rck = \lambda \nu(B)$

\noindent 2) follows from lemma~\ref{lem:pp} :
\begin{eqnarray*}
  2 \gamma(B,D) &=& \bE \lck \lp \Phi(B) - \Phi(D) \rp^2 \rck \\
  &=& \bE \lck \lp \Phi(B \backslash D) + \Phi(B\cap D) - \Phi(D \backslash B) - \Phi(D \cap B) \rp^2 \rck \\
  &=& \bE \lck \lp \Phi(B_D)- \Phi(D_B) \rp^2 \rck \\
  &=& \bE \lck \Phi^2(B_D) \rck +\bE \lck \Phi^2(D_B) \rck -
  2 \bE \lck \Phi^2(B_D) \Phi^2(D_B) \rck \\
  &=& \lambda \nu(B_D) + \lambda^2 \int_{B_D \times B_D} g(x-y) \dd x \dd y
   + \lambda \nu(D_B) + \\
   &&\lambda^2 \int_{D_B \times D_B} g(x-y) \dd x \dd y\\
  && - 2 \lp \lambda \nu(B_D \cap D_B) + \lambda^2 \int_{B_D \times D_B} g(x-y) \dd x \dd y \rp \\
  &=&\lambda \lp \nu(B_D) + \nu(D_B) \rp  + \lambda^2 \lp
   \int_{B_D \times B_D} g(x-y) \dd x \dd y \right. \\
   && \left.+ \int_{D_B \times D_B} g(x-y) \dd x \dd y - 2 \int_{B_D \times D_B} g(x-y) \dd x \dd y \rp
   \end{eqnarray*}

\noindent 3) follows from the approximation
$\bP \lck \lce \Phi(B) = 1 \rce \cap \lce \Phi (D) =1 \rce \rck \approx  \lambda^2 \nu(B) \nu(D) g(r)$ in lemma~\ref{lem:pp}.4).
\finpreuve

%\newpage
\section{Proof of Equations~(\ref{eq:IMSE}) and~(\ref{eq:optimalmesh})}
\label{app:C}

\noindent Let $B$ a square centered at $0$ of area $\nu(B)=b^2$. We denote by $\nabla \lambda(x)$ the gradient vector
$$\nabla \lambda(x) = \nabla \lambda(x_1,x_2) = \lp \partial_1 \lambda(x), \partial_2 \lambda(x) \rp^T = \lp \frac{\partial \lambda(x)}{\partial x_1}, \frac{\partial \lambda(x)}{\partial x_2} \rp^T.$$
\noindent By the following Taylor expansion around the origin
\begin{equation*}
\lambda(x|\Phi_{S_{obs}}) = \lambda(0|\Phi_{S_{obs}}) + x^T \nabla \lambda(0|\Phi_{S_{obs}}) + o(\|x\|),
\end{equation*}
we obtain that:
\begin{eqnarray*}
\bE[\widehat \lambda(x | \Phi_{S_{obs}} )] &=& \frac{\bE[\Phi(B)]}{\nu(B)} = \frac{1}{\nu(B)} \int_B \lambda(x|\Phi_{S_{obs}}) \dd x \\
&\approx& \frac{1}{\nu(B)} \int_B \lambda(0|\Phi_{S_{obs}}) + x^T \nabla \lambda(0|\Phi_{S_{obs}}) \dd x \approx \lambda(0|\Phi_{S_{obs}}) \\
&\approx& \lambda(x|\Phi_{S_{obs}}) - x^T \nabla \lambda(0|\Phi_{S_{obs}}),
\end{eqnarray*}
\noindent and so
\begin{eqnarray*}
&&\int_B \lp \lambda(x|\Phi_{S_{obs}})-\bE[\widehat \lambda(x|\Phi_{S_{obs}})] \rp^2 \dd x \approx \int_B \lp x^T \nabla \lambda(0|\Phi_{S_{obs}}) \rp^2 \dd x \\
&\approx& \int_{-b/2}^{b/2} \int_{-b/2}^{b/2} \lp x_1 \partial_1 \lambda(0|\Phi_{S_{obs}}) + x_2 \partial_2 \lambda(0|\Phi_{S_{obs}}) \rp^2 \dd x_1 \dd x_2 \\
%&\approx& \int_{-b/2}^{b/2} \int_{-b/2}^{b/2} \lp x_1 \partial_1 \lambda(0|\Phi_{S_{obs}}) \rp^2 + \lp x_2 \partial_2 \lambda(0|\Phi_{S_{obs}}) \rp^2
%													+ 2x_1x_2\partial_1\lambda(0|\Phi_{S_{obs}})\partial_2\lambda(0|\Phi_{S_{obs}}) dx_1 dx_2 \\
%&\approx& \int_{-b/2}^{b/2} \int_{-b/2}^{b/2} x_1^2 (\partial_1 \lambda(0|\Phi_{S_{obs}}))^2 dx_1 dx_2
%													+ \int_{-b/2}^{b/2} \int_{-b/2}^{b/2} x_2^2 (\partial_2 \lambda(0|\Phi_{S_{obs}}))^2 dx_1 dx_2 \\
%&+& \int_{-b/2}^{b/2} \int_{-b/2}^{b/2} 2x_1x_2 \partial_1 \lambda(0|\Phi_{S_{obs}}) \partial_2 \lambda(0|\Phi_{S_{obs}}) dx_1 dx_2 \\
%&\approx& \left[ \frac{x_1^3}{3} \right]_{-b/2}^{b/2} (\partial_1 \lambda(0|\Phi_{S_{obs}}))^2 \left[x_2\right]_{-b/2}^{b/2}
%													+ \left[ \frac{x_2^3}{3} \right]_{-b/2}^{b/2} (\partial_2 \lambda(0|\Phi_{S_{obs}}))^2 \left[x_1\right]_{-b/2}^{b/2} \\
&\approx& \frac{b^4}{12} \left[ (\partial_1\lambda(0|\Phi_{S_{obs}}))^2 + (\partial_2 \lambda(0|\Phi_{S_{obs}}))^2 \right]. \\
\end{eqnarray*}
\noindent By consequence, and using Proposition~(\ref{eq:varest}) we have
\begin{eqnarray*}
IMSE \lp \widehat \lambda(x|\Phi_{S_{obs}}) \rp &\approx& \lp \sum_{x_i \in S_{obs}} \int_{B_i} \lp \lambda(x|\Phi_{S_{obs}}) - \bE[\widehat \lambda(x|\Phi_{S_{obs}})] \rp^2 \dd x \rp \\
&&+ \frac{\lambda \nu(S)}{b^2} \\
&\approx& \lp \sum_{x_i \in S_{obs}} \frac{b^4}{12} \| \nabla \lambda(x_i|\Phi_{S_{obs}}) \|^2 \rp + \frac{\lambda \nu(S)}{b^2} \\
&\approx& \frac{b^4}{12} \int_S \| \nabla \lambda(x|\Phi_{S_{obs}}) \|^2 \dd x + \frac{\lambda \nu(S)}{b^2}.
\end{eqnarray*}
Deriving by the variable $b$ gives
\begin{equation*}
\frac{\partial IMSE \lp \widehat \lambda(x|\Phi_{S_{obs}}) \rp}{\partial b} = \frac{2b}{12} \int_S \| \nabla \lambda(x|\Phi_{S_{obs}})\|^2 \dd x - \frac{2\lambda\nu(S)}{b^3}
\end{equation*}
and thus the solution of $\frac{\partial IMSE}{\partial b}=0$ is
\begin{equation*}
\nu_{opt}(B) = \sqrt{\dfrac{12 \lambda \nu(S)}{\int_S \| \nabla \lambda(x | \Phi_{S_{obs}}) \|^2 \dd x}}
\end{equation*}
which is a minimum.

%\newpage
\section{Optimal mesh in practice}
\label{app:simuls}
The optimal mesh of the estimation grid, $\nu_{opt}(B)$, depends on the unknown terms $\lambda$ and $\nabla \lambda(x | \Phi_{S_{obs}})$. In this section, we compare different methods which could be used to compute $\nu_{opt}(B)$ in practice.
We simulated point patterns in the unit square from Thomas point processes with different set of parameters: $(\kappa,\mu) \in \lce (10,50); (22,23); (50,10)\rce$ and $\sigma \in \lce 0.001, 0.0025, 0.005, 0.01, 0.025, 0.05 \rce$.
Then, $\nu_{opt}(B)$ is computed as follows. First we estimate/compute the intensity on a $N \times N$ grid. Second, we deduce its gradient from the rate of change between the estimated intensity and its one-cell translated value. Third, we compute $\nu_{opt}(B)$ and its related grid size (in number of pixels).

We used different methods to estimate the intensity.
 The counting method consists in estimating the local intensity in each pixel $B_i$ by  $\widehat{\lambda}(x_i|\Phi_{S_{obs}}) = \Phi(B_i)/\nu(B_i)$.
The global kernel smoothing method is based on a gaussian kernel estimator with global bandwidth and without border
correction, so $\widehat{\lambda}(x|\Phi_{S_{obs}}) = \sum_{\xi \in\Phi} h^{-2}w(\|x-\xi\|/h)$, where $w$ is the density function of
a standard normal distribution. The $k$-nearest neighbours method is an adaptive nonparametric estimation so that
$\widehat{\lambda}(x|\Phi_{S_{obs}}) = 1/(\pi d_k(x)^2)$, with $d_k(x)$ the distance of $x$ to its $k$-nearest neighbour.
Note that we also compute the theoretical value of the intensity from Equation~(\ref{eq:lambdatheo}).

We considered three $N \times N$-grids, with $N \in \lce 100, 200, 500\rce$. We compared the distributions of the optimal grid sizes obtained from 100 realisations of each process and from the different methods, to the theoretical ones.
We select the method which provides the more accurate results and the less sensitivity to the different scenarii.

\begin{figure}[h]
\centering
\includegraphics[width=.6\linewidth]{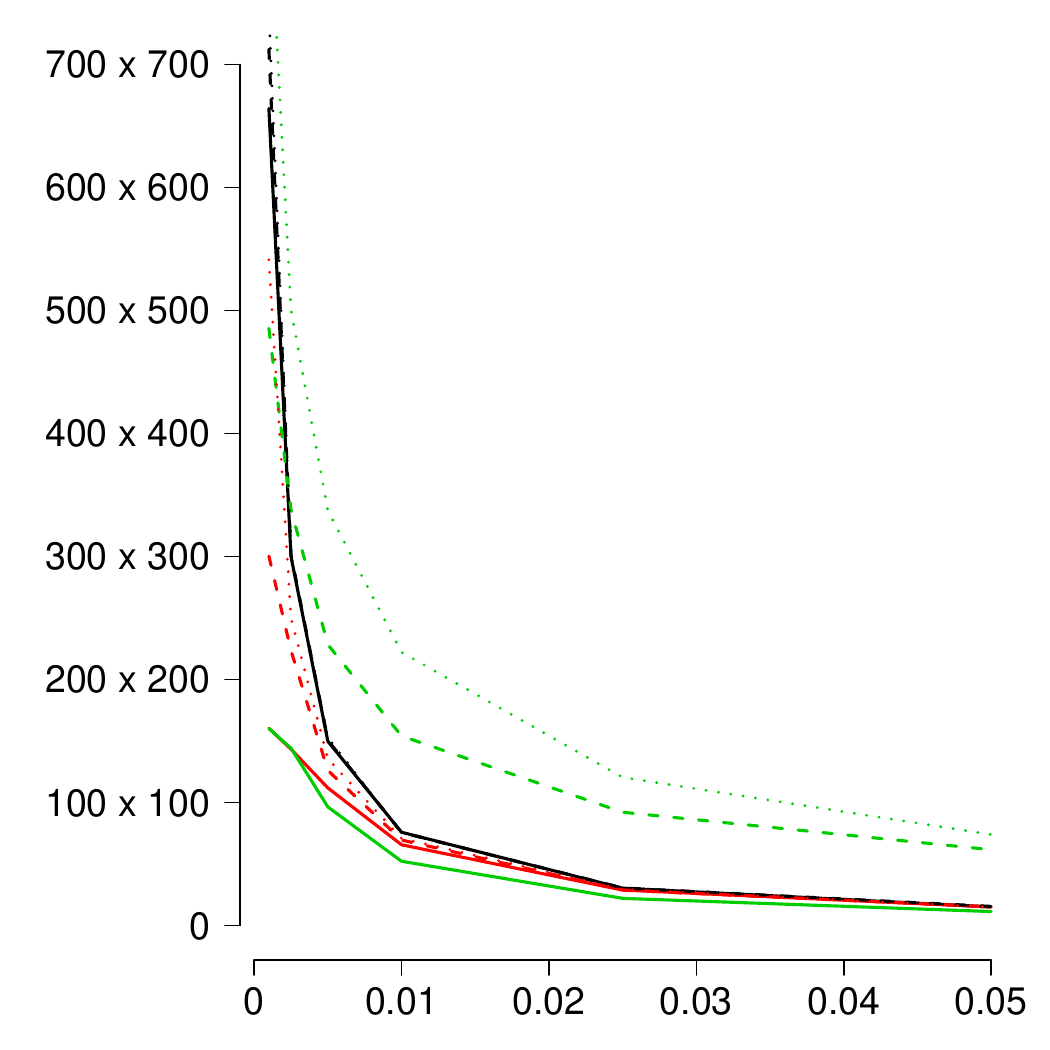}

% {\scriptsize a) \hspace{3.5cm} b) \hspace{3.5cm} c)}

%\includegraphics[width=.33\linewidth]{thoams_kappa10_mu50.pdf}~
%\includegraphics[width=.33\linewidth]{thoams_kappa22_mu23.pdf}~
%\includegraphics[width=.33\linewidth]{thoams_kappa50_mu10.pdf}
\caption{Mean of optimal grid sizes computed from 100 Thomas processes with parameters $\kappa =10$, $\mu=50$ and from the $k$-nearest neighbour based method (green), the kernel based method (red), both evaluated on a $100 \times 100$ grid (solid line), a $200 \times 200$ grid (dashed line) and a $500 \times 500$ grid (dotted line), compared to the theoretical value (black).}
\label{fig:gridopt}
\end{figure}

If the counting method works well when the pattern is strongly aggregated at very
small scale ($\sigma \leq 0.005$), it requires a fine grid and is inaccurate for other scales of clustering. Thus in the following it will be no longer considered.
The other methods provide globally much better values of the optimal grid size.
While $\lambda(x|\Phi_{S_{obs}})$ may be roughly estimated,
the integral of its gradient is sufficiently well approximated to obtain good results.
Figure~\ref{fig:gridopt} shows the optimal grid sizes (in number of pixels) obtained from different size of the estimation grid: $100 \times 100$ (solid line), $200 \times 200$ (dashed line) and $500 \times 500$ (dotted line).
The theoretical grid size is in black and the ones derived from the kernel smoothing and the $k$-nearest neighbours method are in red and green respectively. This figure is related to the Thomas process with parameters $\kappa = 10$, $\mu=50$, and we get similar results from the other set of parameters. It appears that the $k$-nearest neighbours based method is very sensitive to the size of the estimation grid and tends to over-estimate the optimal grid size. The kernel based method under-estimates the optimal grid size when the estimation grid is not fine enough and when the scale of clustering in very small.

From this simulation study, we recommend the kernel based method on
a $200\times200$ grid to first estimate $\nabla \lambda(x|\Phi_{S_{obs}})$ and then compute the optimal mesh or equivalently the optimal grid size.

\end{document}